\documentclass{article}
\usepackage{authblk}
\usepackage{amsmath}
\usepackage{amsfonts}
\usepackage{amssymb}
\usepackage{amscd}
\usepackage{mathtools}

\usepackage{graphicx,color}
\usepackage{fullpage}
\usepackage{tikz}
\usepackage{subcaption}
\usepackage{color,soul}

\newcommand{\norm}[1]{\left\lVert#1\right\rVert}
\newcommand{\bs}{\boldsymbol}
\newcommand{\ft}{\text}
\newcommand{\tb}{\textbf}
\newcommand{\ed}{{\text{d}}}

\title{A mixed variational principle in nonlinear elasticity using Cartan's
moving frame and implementation with finite element exterior calculus}
\author[1]{Bensingh Dhas}
\author[1]{Jamun Kumar N}
\author[1]{Debasish Roy\thanks{royd@iisc.ac.in}}
\author[2]{J N Reddy\thanks{ jnreddy@tamu.edu}}
\affil[1]{Centre of Excellence in Advanced Mechanics of Materials, Indian Institute of Science, Bangalore, India}
\affil[2]{J. Mike Walker’66 Department of Mechanical Engineering, Texas A\&M University, College Station, TX 77843-3123, USA}

\date{}
\begin{document}
\maketitle
\begin{abstract}
    This article offers a new perspective for the mechanics of solids using moving Cartan's frame, specifically discussing a mixed variational principle in non-linear elasticity. We treat quantities defined on the co-tangent bundles of reference and deformed configurations as additional unknowns. Such a treatment invites compatibility of the fields with base-space (configurations of the body), so that the configuration can be realised as a subset of the Euclidean space. We first rewrite the metric and connection using differential forms, which are further utilised to write the deformation gradient and Cauchy-Green deformation tensor in terms of frame and co-frame fields. The geometric understanding of stress as a co-vector valued 2-form fits squarely within our overall program. We show that, for a hyperelastic solid, an equation similar to the Doyle-Erciksen formula may be written for the co-vector part of stress. Using these, we write a mixed energy functional in terms of differential forms, whose extremum leads to the compatibility of deformation, constitutive rules and equations of equilibrium. Finite element exterior calculus is then utilised to construct a finite dimensional approximation for the differential forms appearing in the variational principle. These approximations are then used to construct a discrete functional which is then numerically extremised. This discertization leads to a mixed method as it uses independent approximations for differential forms related to stress and deformation gradient. The mixed variational principle is then specialized for 2D case, whose discrete approximation is applied to problems in nonlinear elasticity. An important feature of our FE technique is the lack of additional stabilization. From the numerical study, it is found that the present discretization also does not suffer form locking and related convergence issues.\\
    
\end{abstract}

\section{Introduction}
Mixed or complementary variational principles have a long history in solid
mechanics \cite{oden1974dual}. These variational principles are of significant
usefulness in developing approximation techniques  where conventional or single
field approximations perform poorly \cite{Veubeke1974,simo1986}. The Hu-Washizu
(HW) principle, for instance, is a three field variational approach commonly
used to construct FE approximations in nonlinear solid mechanics.  The HW
variational principle takes deformation gradient and stress as additional inputs
along with deformation. From a geometric perspective, deformation gradient and
stress are infinitesimal (tangent space based) quantities whose origins can be
traced to the geometric hypothesis placed on the configurations of the body.
However, scarce little has been achieved in understanding these principles from
a geometric standpoint. Electromagnetism is a classical example of a field
theory which was reformulated using ideas from geometry. The theory of
differential forms \cite{hehl2012} played a pivotal role in this reformulation,
leading to a deeper understanding of the field equations and novel numerical
techniques to approximate their solution \cite{bossavit1988}. Later, within the
broad program of geometrisation of physics, differential forms became an
indispensable tool in the mathematical description of physical phenomena. 
Even though linear elasticity has a lot in common with the theory of
electromagnetism \cite{eshelby1951}, the scenario is quite different for
nonlinear elasticity. While electric and magnetic fields are the quantities of
interest in electromagnetism, the interest in elasticity is on stresses and
strains. The distinction however, arises from the nature of the base-spaces
(configurations). In the case of electromagnetism, the base-space is fixed and
only fields (like electric and magnetic fields) have an evolution. In contrast,
for an elastic solid, the base manifold (or configuration) evolves with the
deformation process. This not-so-subtle difference prevents one from mimicking
the techniques used to geometrically formulate electromagnetism in the context
of nonlinear elasticity. Moreover, while electromagnetism can be formulated
entirely using differential forms, nonlinear elasticity requires both vector
fields and differential forms for a complete description. 

The method of moving frames developed by Cartan
\cite{clelland2017,guggenheimer1963} is an effective tool that works seamlessly
with both vector fields and differential forms. Built on the theory of exterior
calculus, Cartan's moving frames can also encode the connection information on a
manifold, which is indispensible for differentiating vector fields on a
manifold. To each point of the configuration, the method of moving frames assign
a bunch of orthonormal vectors called frame fields. The rate at which these
frame fields vary across the configuration defines the connection 1-forms. These
connection 1-forms must however satisfy the structure equations so that the
parallel transport they encode conforms to the underlying manifold structure,
which will presently be assumed Euclidean. Apparently, attaching a set of
vectors to a material point is nothing new in continuum mechanics. Many such
models have been put forth, starting from Cosserat to micro-morphic theories;
these theories are sometimes referred to as micro-continuum theories
\cite{eringen2012}. None of them however encode the connection information of
the deforming body whilst evolving the frames or the directors. For these
models, directors are just additional degrees of freedom to hold energy. A major
difficulty with this point of view is that it does not clarify the geometry
within which the model is working. An immediate consequence is that it is
impossible to give a co-ordinate independent meaning to the derivatives
appearing in the equations of motion. A similar scenario holds, for instance,
with shell theories, including the computational schemes \cite{simo1989} which
are built on models that include directors as degrees of freedom in addition to
the mid-surface deformation. Clearly, these director degrees of freedom are
proxies for the connection induced on the surface. However, since the directors
do not satisfy Cartan's structure equations, it may not be possible to realise
the deformed shell as subsets of an Euclidean space, let alone the more general
case of shells with intrinsic curvature (i.e. defects). 

An attempt at utilizing Cartan's moving frames to formulate the equations of
elasticity  was made by Frankel \cite{frankel2011}; however his  efforts went
largely unnoticed and fell short of offering an appropriate computational
implementation. In this work, we formulate the kinematics of an elastic body
in the language of differential forms. The advantages of having the kinematics
formulated in terms of moving frames are twofold. First, important kinematic
quantities are described using differential forms that explicate on issues related to compatibility. 
Second, the geometric hypothesis behind the kinematics is made explicit. The
hypothesis that the geometry of non-linear elasticity is Euclidean \cite{mfe}
conforms well with the tensor fields that typically describe the local state of
deformation -- the right Cauchy-Green deformation tensor to wit, whose roots can
be traced to the (Euclidean) metric tensor. Moreover, compatibility equations
for both the deformation gradient and Cauchy-Green deformation tensor
\cite{yavari2013} depend on the geometry of the configuration. Compatibility in
terms of the deformation gradient is related to the vanishing of the torsion
tensor while compatibility in terms of the Cauchy-Green deformation tensor is
related to the vanishing of the curvature tensor \cite{blume1989}. Both notions
of compatibility are however related to the affine connection placed on the
configuration. The continuum mechanical definition of stress also has a
geometric meaning; Frankel \cite{frankel2011} describes Cauchy stress as a
bundle valued differential form. The basic idea in his construction is to
decompose the stress tensor into a traction ($1-$form) and an area component
($2-$form). Such a program was later pursued by Segeve and Rodnay
\cite{segev1999} and Kanso \textit{et al.} \cite{kanso2007}. However, both
Segeve and Rodnay and Kanso \textit{et al.} did not pursue a variational
principle using this description. From the decomposition of  Cauchy stress, it
is clear that the constitutive rule needs to be written only for the traction
component since the area component is determined kinematically.

Conventionally, numerical techniques for nonlinear elasticity were largely based
on single field approximations. It was soon realised that these methods suffered
from numerical instabilities, thus affecting the convergence. Displacement based methods with additional stabilization, which barely carries any physical meaning, is a common
technique to circumvent these difficulties. Methods like assumed strain and
enhanced strain techniques belong to this class; these methods introduce
addition terms in the energy function whose origin is purely numerical. Within
nonlinear elasticity, techniques based on mixed FE methods are already the
preferred choice \cite{shojaei2018,shojaei2019,angoshtari2017} for large
deformation problems since they avoid numerical instabilities like locking and
preserve important conserved quantities.  However, constructing stable mixed
finite elements is difficult even in the case of linear
elasticity. A few researchers have turned to ideas from differential geometry to construct stable,
well performing mixed-FE techniques for nonlinear elasticity; we cite Yavari
\cite{yavari2008} as an example of one such attempt where tools from discrete
differential forms \cite{hiraniThesis} were utilised to construct stable
discretisation schemes. These schemes try to preserve important properties of
differential forms even in the discrete setting. And yet, there is barely an attempt at developing a variational principle in solid mechanics to unify the posing and numerical solution of a geometrically conceived model.

Finite element exterior calculus (FEEC) \cite{arnold2006} is an FE technique
developed by Arnold and co-workers to unify finite elements like Raviart-Thomas,
N\'ed\'elec \cite{raviart1977,nedelec1980,nedelec1986} and other carefully
handcrafted elements under a  common umbrella. FEEC relies on the theory of
differential forms to achieve this unification
\cite{bossavit1988,arnold2006,arnold2010,hiptmair1999}. The algebraic and
geometric structures brought forth by differential forms are instrumental in
achieving this. Recently, Angoshtari \textit{et al.} \cite{angoshtari2017} and
Shojaei and Yavari \cite{shojaei2018} have brought on ideas from algebraic
topology to discretize the equations of nonlinear elasticity. These authors have
constructed mixed FE  techniques using HW variational principle. Their
discretization was based on a  differential complex which unfortunately is not the same as 
the de Rham complex. These methods still require
stabilization terms in the three dimensional case \cite{shojaei2019}. Moreover, from the
description of the complex given in Angoshtari \textit{et al.}, it is not clear
how the HW variational principle is related to the complex and how the operators
defining the complex are affected by the connection placed on the configuration.

The goals of this article are twofold. The first is to develop a mixed
variational principle for nonlinear elasticity that takes differential forms as
its input argument. Towards this, we first reformulate the kinematics and
kinetics of an elastic solid using differential forms. The kinematics of
deformation is laid out via Cartan's method of moving frames. The structure
equations associated with the moving frames establish the important relationship
between the geometric hypothesis of the configuration and measures of
deformation. The geometric understanding of stress as a bundle valued
differential form is then exploited to write the kinetics in terms of
differential forms. These two ideas are then used to rewrite the conventional HW
variational principle, which now has a bunch of differential forms and
deformation as its input arguments. The proposed mixed functional is then
extremised with respect to these differential forms to arrive at the equations
of mechanical equilibrium, constitutive relation and compatibility constraints.
The second goal of this work is to use FEEC to discretize the proposed mixed
variational principle. Towards this end, the spaces $\mathcal{P}_r\Lambda^1$ and
$\mathcal{P}_r^{-}\Lambda^1$ are used to discretize the differential forms
describing the kinetics and kinematics. Using these approximations, a discrete
mixed functional is arrived at, which can then be extremized using numerical
optimization techniques. 


The rest of the article is organized as follows. A brief introduction to
Cartan's moving frames and the associated structure equations are presented in
Section \ref{sec:movingFrame}. The  kinematics of an elastic body is then
reformulated in Section \ref{sec:kinematics}, using the idea of moving frames.
In this section, important kinematic quantities like deformation gradient and
right Cauchy-Green deformation tensor are described using frame and co-frame
fields. This section also contains a discussion on affine-connections using
connection 1-forms. In Section \ref{sec:stress}, we introduce stress as a
co-vector valued differential 2-form; this interpretation is originally due to
Frankel \cite{frankel2011}. We then derive a relationship similar to the
Doyle-Ericksen formula relating the stored energy density function to the
traction $1-$form. In Section \ref{sec:variational}, we rewrite the mixed
variational principle in terms of differential forms describing the kinematics
and kinetics of motion. We then show that variation of the mixed functional with
respect to different input arguments leads to the compatibility of deformation,
constitutive rule and equations of equilibrium. We also remark on the
interpretation of stress as a Lagrange multiplier enforcing compatibility of
deformation. Section \ref{sec:discreteForms} discusses a discrete  approximation
of differential forms on a simplicial manifold. While these ideas have their
roots in the work of Whitney \cite{whitney2012}, we adopt a description of the
FEEC within the framework of Cartan's moving frames. These are utilized to
construct a discrete approximation to the mixed variational principle, which is
numerically extremized using  Newton's method. This FE approximation is then
applied to standard benchmark problems in nonlinear elasticity in order to
assess the performance of the numerical technique against instabilities like
volume and bending locking. Finally, in Section \ref{sec:conclusion}, we discuss
on the usefulness of the moving frames in formulating other theories in
nonlinear solid mechanics (like Kirchhoff shells and dislocation mechanics) and
the extension of the present numerical techniques to such nonlinear theories.

\subsection{A remark on notations}
We do not use bold face letters to distinguish between a scalar, a vector or a
tensor. Indices are used to index objects of the same kind and not the
components; for example, if we have three $1-$forms, we may denote the $i$th
$1-$form by $\theta^i$. A symbol with one index does not mean that it is a
component of a vector or a $1$-form. The objects featured in the theory are
defined wherever they appear first. Often in nonlinear elasticity, lower and
upper case indices are used to distinguish objects in the reference and deformed
configurations. We do not follow this convention since we adopt separate
notations for the same object defined in the reference and deformed
configurations. We follow the convention of Einstein summation over repeated indices. In places where the summation convention is not followed, we make it explicit.

\section{Cartan's moving frame}
\label{sec:movingFrame}
In this section, we present a brief introduction to the method of moving frames;
our motive being to review some  basic results so that the kinematics of an
elastic solid can be written in terms of moving frames. For a detailed
exposition on moving frames, the reader may consult \cite{clelland2017} and
\cite{guggenheimer1963}. Cartan introduced the method of moving frames as a tool
to study the geometry of surfaces. These techniques were later extended to study
the geometry of abstract manifolds. Common examples of moving frames include the
Frenet frame for a curve and Durboux frame for a surface.  

We denote the reference and deformed configurations of a body by $\mathcal{B}$
and $\mathcal{S}$; the respective tangent bundles are denoted by $T\mathcal{B}$ and
$T\mathcal{S}$. Both $\mathcal{B}$ and $\mathcal{S}$ are smooth
manifolds with boundaries; their boundaries are denoted by $\partial \mathcal{B}$
and $\partial \mathcal{S}$. These configurations are endowed with a $C^\infty$
chart from which they inherit their smoothness. Positions (placements) of a
material point in the reference and deformed configurations are denoted by $X$
and $x$ respectively.

At each tangent space of a configuration, we choose a collection of orthogonal
vectors, which we call the frame. The orthogonality of the frame
field is with respect to the Euclidean inner product of the respective
configuration. The frame fields of the reference and deformed configurations are
denoted by $E_i$ and $e_i$ respectively. A collection of frame fields that span
the tangent spaces constitutes a moving frame or simply a frame (allowing for a slight abuse of the terminology). We denote frames
for the reference and deformed configurations  by
$\mathcal{F}_\mathcal{B}=\{E_1,...,E_n\}$ and
$\mathcal{F}_\mathcal{S}=\{e_1,...,e_n\}$. Given a frame for a tangent bundle,
the natural (algebraic) duality between tangent and co-tangent spaces induces a
co-frame for the co-tangent bundle as well. These co-frames (at a point)
constitute a basis for the co-tangent spaces of the respective configurations.
We denote the co-frames of the reference and deformed configurations  by
$\mathcal{F}^*_\mathcal{B}=\{E^i,...,E^n\}$ and
$\mathcal{F}^*_\mathcal{S}=\{e^1,...,e^n\}$ respectively, where $E^i$ and $e^i$
are sections from the cotangent bundles of the reference and deformed
configurations. The natural duality between frame and co-frame fields of the
reference and deformed configurations may be written as,
\begin{equation}
    E^i(E_j)=\delta^i_j;\quad
    e^i(e_j)=\delta^i_j; \quad
    E^i\in T^*\mathcal{B}, e^i\in T^*\mathcal{S}
\end{equation}
For a material point in the reference
configuration, the differential of position is denoted by $\ed X$. In terms of the frame and co-frame fields, it can be written as,
\begin{equation}
    \ed X= E_i\otimes E^i
    \label{eq:refDiff}
\end{equation}
Similarly, in terms of the frame and co-frame fields in the deformed configuration,
 the differential of position is given by,
\begin{equation}
    \ed x= e_i\otimes e^i
    \label{eq:defDiff}
\end{equation}
From the definition of $\ed X$ \eqref{eq:refDiff}, it follows that a tangent
vectors from the reference configuration is  mapped to itself under $\ed X$. To
see this, choose $V\in T_X \mathcal{B}$  with $V=c^i E_i$. Substituting the
latter and using the definition of $\ed X$, we arrive at  $\ed X(V)=c^j E_i
E^i(E_j)$. Using the duality between the frame and co-frame fields, we conclude that
$\ed X(V)=V$.  Similarly, $\ed x$ maps a tangent vector from the deformed
configuration to itself.  The differential of a position vector in the
reference or deformed configuration is thus an identity map on the corresponding
tangent space.

Similar to the differential of position, one may also define the differential of
a frame, which in the reference configuration is given by,
\begin{equation}
    \ed E_i=\gamma^j_i\otimes E_j
    \label{eq:differentialFrame}
\end{equation}
where, $\gamma^i_j$ is called the connection matrix; it contains 1-forms as its
entries. Because of the orthogonality between the frame fields, the connection
matrix is skew symmetric, i.e. $\gamma^i_j=-\gamma^j_i$. Similarly, the differential of a frame for the deformed configuration is given by,
\begin{equation}
    \ed e_i=\bar{\omega}^j_i\otimes e_j
    \label{eq:differentialFrameDeformed}
\end{equation}
$\bar{\omega}^i_j$ is the connection matrix of the deformed frame fields. It is also skew, i.e. $\bar{\omega}^i_j=-\bar{\omega}^j_i$.

For a given choice of connection 1-forms and co-frame fields, there are certain
compatibility conditions (Poincar\'e relations) which guarantee the existence of the
placements $X$ and $x$.  These equations are called Cartan's structure
equations. In the present context (of all manifolds being Euclidean), the first compatibility condition establishes the torsion free nature
of the configuration.  For the reference and deformed configurations, this
condition may be written as,
\begin{equation}
    \ed^2{X}=0;\;\ed^2 x=0
\end{equation}
Plugging \eqref{eq:refDiff} and \eqref{eq:defDiff} into the above equation and making use of \eqref{eq:differentialFrame} and \eqref{eq:differentialFrameDeformed} leads to,
\begin{equation}
    \ed E^i=\gamma^i_j \wedge E^j;\;
    \ed e^i=\bar{\omega}^i_j \wedge e^j
    \label{eq:torsionFree}
\end{equation}
The second compatibility condition presently establishes that the reference and deformed
configurations are curvature-free. This leads to the following conditions on the
reference and deformed frame fields,
\begin{equation}
    \ed^2 E_i =0;\;\; \ed^2e_i=0
\end{equation}
Using the differential of the frame fields in the above equations yields,
\begin{equation}
    \ed \gamma^i_j= \gamma^i_k\wedge\gamma^k_j;\;\;
    \ed \bar{\omega}^i_j= \bar \omega^i_k\wedge\bar \omega^k_j
    \label{eq:curvatureFree}
\end{equation}
For a simply connected body, the structure equations \eqref{eq:torsionFree} and
\eqref{eq:curvatureFree} (for the reference and deformed configurations) provide
the necessary kinematic closure to ensure that the configurations can be
embedded with an Euclidean space. Indeed, without this closure effected by the
structure equations, a model cannot in general produce a deformed configuration
which is a subset of an Euclidean space.

\subsection{Differentials of position and frame}
\begin{figure}
    \centering
    \includegraphics[scale=.95]{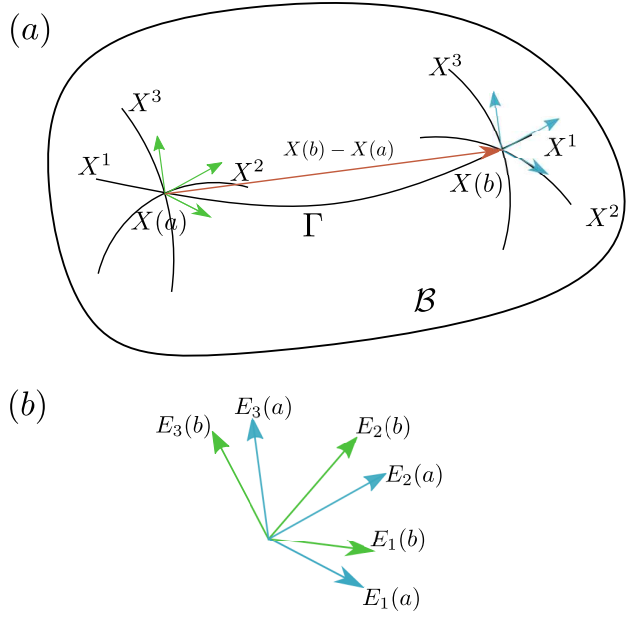}
    \caption{The coordinate lines and the frame field generated from these
        coordinate lines are shown in (a). The frame fields at $X(a)$ and
        $X(b)$ are shown in (b); we have moved the frames to the same point so that
        it is convenient to interpret the change. We have used the
        notation $E_i(a)$ and $E_i(b)$ to indicate the frame at the material
        points $X(a)$ and $X(b)$.}
    \label{Fig:cordinateSystem}
\end{figure}

We now present the geometric meaning of the infinitesimal quantities
(differentials of position and frame) introduced in the last subsection.
Consider the differential of the position for the reference configuration given
in \eqref{eq:refDiff}. For a given coordinate system, the position  $X$ is a
smooth function of its coordinates $(X^1,X^2,X^3)$.  Let $\Gamma$ be a curve
parametrized by its arc length, $\Gamma:[a,b]\rightarrow \mathcal{B}$,
$G\left(\frac{d\Gamma}{ds},\frac{d\Gamma}{ds}\right)=1$, where $G(.,.)$ denotes
the metric tensor of the reference configuration. 
The frame fields $E_i$ can be constructed by an orthonormalisation (Gram-Schmidt
procedure) of the tangent vectors to the coordinate curves. Fig.
\ref{Fig:cordinateSystem} shows the coordinate curves and frame at $X(a)$ and
$X(b)$. The tangent vector to the curve $\Gamma$ is written as
$\frac{d\Gamma}{ds}=c^iE_i$, where $c^i$ are real numbers. Now, the differential
of position, which is a vector valued 1-form, can be integrated along the curve
$\Gamma$ to produce a vector; this  vector translates the position $X(a)$ to  
$X(b)$. This translation may be formally written as,
\begin{align}
    \nonumber
    X(b)-X(a) & =\int_a^b \ft{d}X(c^iE_i)d s       \\
    \nonumber
                        & =\int_a^b c^iE_j E^j(E_i)d s, \\
                        \nonumber
                        & =\int_a^b c^iE_j \delta^j_i d s, \\
                        & =\int_a^b c^iE_i d s.
\end{align}
In the last equation, $E_i$ and $c^i$ can vary along the curve $\Gamma$. The
above interpretation of $\ed X$ is very similar to that of 1-forms as real
numbers defined on curves.

We now consider the differential of frame. Integrating \eqref{eq:differentialFrame} along
$\Gamma$, we have,
\begin{align}
    \nonumber
    \int_{a}^b \ed E_i & =\int_a^b\ed E_i\left(\frac{d\Gamma}{ds}\right)\ed s.
\end{align}
Evaluating $\gamma^i_j$ on the tangent vector produces a skew symmetric matrix with real co-efficients. In other words, the above integration can be written as a solution to the ordinary differential equation,
\begin{equation}
    \dot{E}_i=\gamma^j_iE_j,
    \label{eq:ODEFrame}
\end{equation}
$\dot{(.)}$ may be understood as  the derivative with respect to the parameter $s$. Alternatively, one can understand \ref{eq:ODEFrame}, as a restriction of \eqref{eq:differentialFrame} to a curve. Given the initial $E_j(a)$, the solution to \eqref{eq:ODEFrame} is a rotation matrix $R$ which relates the frames between two points on the curve $\Gamma$. This is formally written as,
\begin{equation}
    E_i(s)=R(s) E_i(a).
\end{equation}
Figure \ref{Fig:cordinateSystem} (b) depicts this idea by overlaying the frames at $a$ and $b$. From this discussion, it is clear that the vector
translating the point $X(a)$ to $X(b)$ is dependent on the frame, the co-frame and the curve chosen for integration. However, the vector $X(b)-X(a)$ is path and frame independent since the body $\mathcal{B}$ is a subset of an Euclidean space. This path independence is exactly what the structure
equations enforce.

\subsection{Affine connection via frame fields}
We now discuss the affine connection and covariant differentiation encoded by the connection 1-forms discussed in the previous subsection. An affine connection on a smooth manifold is a device used to differentiate
sections of vector and tensor bundles in a co-ordinate independent manner. Let
$w=\sum_{i=1}^nw^ie_i$ be an arbitrary section from $T\mathcal{S}$. The
covariant derivative of $w$ in the direction of $e_i$ is given by,
\begin{equation}
    \nabla_{e_i} w= \ed w^j(e_i)e_j+w^j\bar{\omega}^k_j(e_i)e_k
\end{equation}
It is easy to check that the above definition is a differential satisfying the properties,
\begin{align}
    \nonumber
    \nabla_{fe_i}w    & =f\nabla_{e_i}w;\;\; f\in \Lambda^0         \\
    \label{eq:covProp}
    \nabla_{e_i}(w+v) & =\nabla_{e_i}w+\nabla_{e_i}v; \; w,v \in TS \\
    \nonumber
    \nabla_{e_i}(fw)  & =e_i[f]w+f\nabla_{e_i}w
\end{align}
Using these properties, it is possible to extend the above definition of covariant differentiation to arbitrary tensor fields \cite{mfe}.

\section{Kinematics}\label{sec:kinematics}
The deformation  map  sends the placement of material points in the reference
configuration to their corresponding placements in the deformed configuration.
We denote the deformation map by $\varphi$ so that $x=\varphi(X)$. The
differential of the deformation or the deformation gradient, denoted by $\ed
\varphi$, maps the tangent space of the reference configuration to the
corresponding tangent space in the deformed configuration.  For an assumed frame
field (for both reference and deformed configurations), the differential of
deformation can be obtained by pulling the co-vector part of the
deformed configuration's differential of position back to the reference configuration.
\begin{align}
    \nonumber
    \ft{d} \varphi& =e_i\otimes \varphi^*(e^i)                                                       \\
                  & =e_i\otimes\theta^i; \quad \theta^i \in T^*\mathcal{B},e_i\in\mathcal{F}_\mathcal{S}
    \label{eq:deformationOneForm}
\end{align}
In writing \eqref{eq:deformationOneForm}, we have introduced the following
definition: $\theta^i:=\varphi^*(e^i)$.  In our construction, the 1-forms
$\theta^i$ contains local information about the deformation map $\varphi$. The
1-forms $\theta^i$ are a  primitive variable in our theory; we call these
differential forms, the deformation 1-forms.  From \eqref{eq:deformationOneForm},
we see that the vector leg of the deformation gradient is from the deformed
configuration, while the co-vector leg is from the reference configuration,
making it a two-point tensor \cite{mfe}. The action of the deformation gradient
on a vector $V(X)\in T_{X}\mathcal{B}$ is given by,
\begin{equation}
    \ed \varphi(V)=e_i \theta^i(V)
\end{equation}
Since $\theta^i(V)$ are real numbers, the above equation is a linear combination
of tangent vectors from the deformed configuration. Conventionally, deformation
gradient is introduced as the differential of deformation; we have not taken
this perspective since our interest is in constructing mixed variational
principles for nonlinear elasticity. Another important aspect of the above
construction is that, we have written the deformation gradient using a frame and
a co-frame. Contrast this with a conventional mixed method, where the
deformation gradient is identified with its components in a particular
coordinate system.

\subsection{Strain and deformation measures}
The notion of length is central to continuum mechanics; important kinematic
quantities like strain and rate of deformation are derived from it. Indeed, it
may not be possible to assess the state of deformation without a metric
structure (notions of length and angle) for both reference and deformed
configurations. The notion of length is encoded by a  symmetric and positive
definite tensor, defined on the tangent space of the respective configuration.
The metric tensor of the reference and deformed configurations are denoted by
$G$ and $g$ respectively; $G:T_X\mathcal{B}\times
T_X\mathcal{B}\rightarrow\mathbb{R}$ and $g:T_x\mathcal{S}\times
T_x\mathcal{S}\rightarrow\mathbb{R}$.  In this work, we assume the metric
structures  of both reference and deformed configurations to be Euclidean. In
terms of the co-frame field, the metric tensor of the reference configuration is
given as,
\begin{align}
    \nonumber
    G & =\ed X. \ed X\\
        &=(E^j\otimes E_j).(E_i\otimes E^i)\\
      & =E^i\otimes E^i.
\end{align}
The dot product introduced in the above equation is the inner product between the vector legs of $\ed X$, which is computed using the Euclidean inner product.
 Similarly, the metric
tensor in the deformed configurations may be written as,
\begin{align}
    \nonumber
    g & =\ed x. \ed x\\
    &=(e^i\otimes e_j).(e_j\otimes e^i) \\
      & =e^i\otimes e^i.
\end{align}
In terms of the frame fields, the inverses of the metric tensors for the
reference and deformed configurations are written as,
\begin{equation}
    G^{-1}=E_i\otimes E_i;\;\;\;g^{-1}=e_i\otimes e_i.
\end{equation}
Now, the right Cauchy-Green deformation tensor may be  obtained as the pull-back of
the deformed configuration's metric tensor. In terms of the deformation 1-forms,
this relationship may be written as,
\begin{align}
    \nonumber
    C & =\varphi^*(g)                       \\
    \nonumber
      & =\varphi^*(e^i\otimes e^i) \\
      & =\theta^i\otimes\theta^i.
    \label{eq:rightCGMetric}
\end{align}
An alternative way to compute the $C$ is to use the usual definition in continuum
mechanics, $C=\ft{d}\varphi^t\ft{d}\varphi$. Here, $(.)^t$ is understood to be
the adjoint map induced by the metric structure. Using the orthonormality of the
frame field we arrive at,
\begin{align}
    \nonumber
    C & =(\theta^i\otimes e^i) (e_j\otimes \theta^j) \\
      & =\theta^i\otimes\theta^i.
    \label{eq:rightCGDefGrad}
\end{align}
The calculations leading to \eqref{eq:rightCGMetric} and
\eqref{eq:rightCGDefGrad} are exactly the same; only the sequence in which
pull-back and inner product are computed differs.  The Green-Lagrangian strain
tensor may now be written as,
\begin{align}
    \nonumber
    E & =\frac{1}{2}(C-G)                                                     \\
      & =\frac{1}{2}[(\theta^i\otimes\theta^i)-(E^i\otimes E^i)].
\end{align}
The first invariant of the right Cauchy-Green tensor is given by,
\begin{equation}
    I_1=\langle \theta^i,\theta^i \rangle_{G}.
\end{equation}
Here $\langle.,.\rangle_G$ denotes the inner product induced
by $G$. The area forms induced by the co-frame of the
reference configuration are given by,
\begin{equation}
     A^1=E^2\wedge E^3;\quad
     A^2=E^3\wedge E^1;\quad
     A^3=E^1\wedge E^2,\quad
\end{equation}
 Similarly, the area-forms induced by the co-frame of the deformed
configuration are given by,
\begin{equation}
     a^1=e^2\wedge e^3;\quad
     a^2=e^3\wedge e^1;\quad
     a^3=e^1\wedge e^2,\quad
\end{equation}
These area forms $A^i$ and $a^i$ serve as a basis for the space of 2-forms
defined on their respective configurations. The area forms in the deformed configuration
may be pulled back to the reference configuration under the
deformation map. These pulled-back area forms are denoted by
$\mathsf{A}^i:=\varphi^*(a^i)$. In terms of the deformation 1-forms, the pulled
back area forms can be written as,
\begin{equation}
    \mathsf{A}^1=\theta^2\wedge \theta^3;\quad
    \mathsf{A}^2=\theta^3\wedge \theta^1;\quad
    \mathsf{A}^3=\theta^1\wedge \theta^2.
    \label{eq:pullBackAreaForm}
\end{equation}
In terms of the pulled-back area forms, the second invariant of $C$
may now be written as,
\begin{equation}
    I_2=\langle \mathsf{A}^i, \mathsf{A}^i \rangle_G
\end{equation}
In terms of the co-frame fields, the volume forms of reference and deformed
configurations may be written as,
\begin{equation}
     V=E^1\wedge E^2 \wedge E^3;\quad
    v=e^1\wedge e^2 \wedge e^3.
\end{equation}
The pull back of the volume form in the deformed configuration to the reference
configuration is denoted by $\mathsf{V}:=\varphi^*(v)$. In terms of the
deformation 1-forms, $\mathsf{V}$ may be written as,
\begin{equation}
     \mathsf{V}=\theta^1\wedge\theta^2\wedge \theta^3
\end{equation}
Finally, in terms of the  pulled-back volume-form, the third invariant of $C$ is given by,
\begin{equation}
    I_3=({}_{\star} \mathsf{V})^2
\end{equation}
In the above equation ${}_{\star}(.)$ denotes the Hodge star operator, which establishes an isomorphism between the space of $0-$forms and $3-$forms.  We also define $J:=\sqrt{I_3}$ or simply $J={}_{\star} \mathsf{V}$.
\section{Stress as co-vector valued two-form} 
\label{sec:stress} 
In the previous section, we reformulated the  deformation gradient and right
Cauchy-Green deformation tensor in terms of differential forms.
We now present a geometric approach to stress, originally due to Frankel
\cite{frankel2011} and subsequently developed by Segeve and Guy \cite{segev1999}
and Kanso \textit{et al.} \cite{kanso2007}. Even though this approach is
intuitive and geometric, it was never used to construct a variational principle.
This geometric approach to stress has its origins in classical dynamics
\cite{marsden2013}, where force is understood as a co-vector. Identifying force
with a co-vector permits us to write power as a pairing between force and
velocity (action of a 1-form on a vector) without the use of a metric tensor.
Extending this concept  to stress in continuum mechanics is non-trivial and
requires the machinery of bundle valued differential forms \cite{kanso2007}. As
in classical dynamics, this interpretation of stress as a bundle valued
differential form permits us to write the power expended by stress upon
deformation without using a metric (i.e. without explicitly bringing into play the geometric structure of the configuration).

We denote the Cauchy stress tensor by $\sigma$. The traction  acting on an
infinitesimal area element with unit normal $\hat{n}$ is denoted by $t$, which is given by the well
known formula $t=\sigma \hat{n}$. The traction $t$ is a force which depends on the
material point and the area sustaining it. The relation between normal and
traction is postulated to be linear. In the language of differential forms, an
infinitesimal area is regarded as a $2-$form, while from classical dynamics we
also know that force is a co-vector or a $1-$form. Putting these two ideas
together, we are led to a geometric definition of Cauchy stress given by,
\begin{equation}
    \sigma=t^i\otimes a^i;
    \label{eq:Cauchy3Tensor}
\end{equation}
Recall that $a^i$ is the area form of the deformed configuration,
which sustains the traction vector $t^i$. The tensor product in the above
equation is due to the linearity between traction and area forms. From this
equation, it is easy to see that the area form changes orientation if the order of
the co-vectors in the area form is reversed. Geometrically, if there are $m$
linearly independent area forms on the manifold, the stress tensor assigns to
each area form a 1-form called traction. With this understanding, Cauchy stress
may now be identified with a section from the tensor bundle
$\Lambda^1\otimes\Lambda^2$ which has the deformed configuration as its base
space.

In contemporary continuum mechanics, Nanson's formula describes the transformation of an
infinitesimal area under the deformation map \cite{ogden1997}. Geometrically,
Nanson's formula is nothing but the pull-back of an area form in the deformed configuration
under the deformation map. These pulled-back area forms
are given in \eqref{eq:pullBackAreaForm}.
\begin{figure}
    \centering
    \includegraphics[scale=0.5]{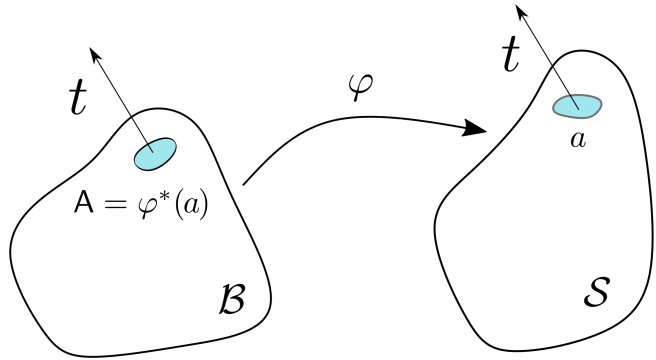}
    \label{Fig:piolaTransform}
    \caption{For an infinitesimal area in the deformed configuration sustaining a traction $t$, Piola transform applies a pull-back on the area leg, while the traction remains unaltered.}
\end{figure}
The first Piola stress may now be obtained by pulling back the area leg of
the Cauchy stress under the deformation map. This partial pull-back of the
Cauchy stress is termed the Piola transform. This relationship may be formally
written as,
\begin{align}
    \nonumber
    P & =t^i\otimes \varphi^*(a^i) \\
      & =t^i\otimes  \mathsf A^i
    \label{eq:Piola3tensor}
\end{align}
Note that in the definition of Piola transform, traction 1-form is left
untouched.  Thus, contrary to convention, Cauchy and first
Piola stresses are now identified as third order tensors, not the usual second
order. This ambiguity can be removed if one applies the Hodge star on the area leg of these two stresses.  In three dimensions, the Hodge star in question establishes an isomorphism between differential forms of degree 2 and 1. The usual definition of stress may thus be recovered as,
\begin{align}
    \label{eq:Cauchy2Tensor}
    \sigma & =t^i\otimes {}_{\star}(a^i) \\
    \label{eq:Piola2Tensor}
    P      & =t^i\otimes {}_{\star}(\mathsf A^i).
\end{align}
Kanso \textit{et al.} made a distinction between the stress tensors given in
\eqref{eq:Cauchy3Tensor}, \eqref{eq:Piola3tensor} and \eqref{eq:Cauchy2Tensor},
\eqref{eq:Piola2Tensor}. However we do not see a need for it, since both the
usual and geometric definitions of stress contain exactly the same
information; only the ranks of these tensors are different.

\subsection{Traction 1-form via stored energy function}
The Doyle-Ericksen formula is an important result in continuum mechanics
\cite{doyle1956nonlinear}, which relates the Cauchy stress and the metric tensor
of the deformed configuration. For a stored energy density function $W$,
Doyle-Ericksen formula gives us the following relationship,
\begin{equation}
    \sigma=2\frac{\partial W}{\partial g}.
    \label{eq:doyleErichsen}
\end{equation} 
In writing \eqref{eq:doyleErichsen}, we have assumed that $W$ is
frame-invariant. From the discussion presented so far, it
is seen that the area leg of the Cauchy stress tensor is determined by the
choice of coordinate system (frame and co-frame fields) for the tangent bundle
of the deformed configuration.  On the other hand, the area leg of the first
Piola stress is determined by both the deformation map and  the co-ordinate system for the tangent bundle
of the deformed configuration. Clearly, the area leg of
a stress tensor does not require a constitutive rule; it is only the traction
component that demands a constitutive rule.

We now claim that for a stored energy function $W$, the traction $1-$form has
the following constitutive rule, 
\begin{equation}
    \label{eq:ConstitutionTraction}
    t^i=\frac{1}{J}\frac{\partial W}{\partial e_i}
\end{equation}
The last equation is in the same spirit as the Doyle-Ericksen formula. To
establish the result in \eqref{eq:ConstitutionTraction}, we first compute
the directional derivative of $W$ along $e_i$,
\begin{subequations}
\label{eq:tiConjei}
\begin{align}
    \label{eq:chainRule}
    \frac{\partial W}{\partial e_i}&=\frac{\partial W}{\partial (\ed \varphi)}
    \frac{\partial (\ed \varphi)}{\partial e_i}\\
    \label{eq:piolaTwoTensor}
    &=(t^j\otimes {}_{\star} \mathsf A^k)(e^i\otimes e_j \otimes \theta^k)\\
    \label{eq:tiConjei-5}
    &=\langle {}_{\star} \mathsf{A}^j, \theta^j \rangle t^i\\
    &=({}_{\star} \mathsf{V})t^i \\
    &=Jt^i
\end{align}
\end{subequations}
We used a chain rule to arrive at the right hand side of \eqref{eq:chainRule}.  In
\eqref{eq:piolaTwoTensor}, the expression for Piola stress (as a two tensor) in
terms of $W$ and the directional derivative of $\ed \varphi$ along $e_i$ are
used to get the right hand side and performing the required contractions lead to
\eqref{eq:tiConjei-5}. The claim is finally established by using the definitions
of pull-back and Hodge star for volume forms.  In three dimensions, constitutive
relations have to be supplied to the three traction $1-$forms. From these
calculation, it is found that the traction 1-forms are conjugate to the frame
fields. 

If the Cauchy stress (the usual definition) generated by a stored energy function is
known, then the expression for the traction 1-form can be computed using the simple
relation $t^i=\sigma^{ij} \hat{n}_i$, where the vector fields $\hat{n}_i$
are chosen to be elements from the frame of the deformed configuration.

\section{Mixed variational principle}
\label{sec:variational}
We first present the conventional HW variational principle for a finitely
deforming elastic body. As mentioned, the HW variational principle takes the
deformation gradient, first Piola stress and deformation map as input arguments.
In the reference configuration, the HW functional for a non-linear elastic solid
can be written as,
\begin{equation}
    \label{eq:conventionalHW}
    I_{HW}=\int_{\mathcal{B}}[W(C)-P:(F-\ed \varphi)]d V- \int_{\partial
    \mathcal{B}}\langle t,\varphi \rangle d A.
\end{equation}
In the above equation, $t=PN$ is the traction defined on the surface $\partial
\mathcal{B}$, whose unit normal is $N$. The integration in the above equation is
with respect to the volume and area forms of the reference configuration. In
\eqref{eq:conventionalHW}, the deformation gradient is assumed to be independent;
this tensor field is denoted by $F$. On the other hand, the deformation gradient computed as
the differential of deformation is denoted by $\ed \varphi$. It is worthwhile to note 
that the second term in the above equation is bilinear in
the Piola stress and the deformation gradient. The variation of the HW functional with
respect to deformation, deformation gradient and first Piola stress leads to the
equilibrium equation, constitutive rule and compatibility of deformation
gradient. This form of HW variational principle has been previously exploited
to formulate numerical solution procedures for non-linear problems in
elasticity; see\cite{angoshtari2017,shojaei2019,shojaei2018}.

\subsection{Mixed variational principle with geometric definitions of
stress and deformation}
We now use the definitions of Cauchy and Piola stresses given in
\eqref{eq:Cauchy3Tensor} and \eqref{eq:Piola3tensor} respectively to
rewrite the HW variational principle such that it takes deformation $1-$forms,
traction $1-$forms and deformation map as inputs. We also assume that
compatible frames for the reference and deformed configurations are given. This
assumption permits us to eliminate the frame fields from the list of unknowns.
The mixed functional may be now written as,
\begin{equation}
    I(\theta^i,t^i,\varphi)=\int_{\mathcal{B}}W(\theta^i)\ed V-(t^i\otimes
     \mathsf{A}^i)\dot{\wedge}(e_i \otimes (\theta^i- \ft{d}\varphi^i))
    -\int_{\partial \mathcal{B}}\langle t^\sharp,\varphi \rangle dA.
    \label{eq:modifiedHW0}
\end{equation}
In \eqref{eq:modifiedHW0}, $\dot{\wedge}$ denotes a bilinear map. For $\alpha
\in T^*\mathcal{S}$, $v\in T\mathcal{S}$ and $a,b \in \Lambda^n(\mathcal{B})$,
$n\geq 1$, the action of this map is given by $(\alpha\otimes
a)\dot{\wedge}(v\otimes b)=\alpha(v)a\wedge b$. Note that the definition of
$\dot{\wedge}$ given here is a little different from the one in Kanso \textit{et
al.} \cite{kanso2007}. Specifically, we do not use the metric tensor. From our
definition of $\dot{\wedge}$, it is seen that the work done by stress on
deformation is metric independent. This property of our current variational
formulation brings the continuum mechanical definition of stress a step closer
to the definition of force (as a $1-$form) in classical mechanics
\cite{marsden2013}. We did not write the volume form in the second term on the
RHS of \eqref{eq:modifiedHW0}, since the outcome of $\dot{\wedge}$ is a 3-form
which can be integrated over the reference configuration to produce work done by
traction 1-forms on deformation.  Also note that the second term on the RHS
in \eqref{eq:modifiedHW0} is equivalent to the second term in
\eqref{eq:conventionalHW}; however now the relationship between the different
arguments is multi-linear. The functional given in \eqref{eq:modifiedHW0} can also be discussed within the framework given by Oden and Reddy \cite{oden1974dual} for the construction of complementary variational principles.

\textit{Remark 1:} In writing \eqref{eq:modifiedHW0}, we have postulated that the
geometry of the body is Euclidean and it is frozen during the deformation
process. Indeed, within the present set-up, this assumption can be relaxed by permitting
non-integrability in the connection and deformation 1-forms (i.e. by
incorporating source terms in the structure equations).

\textit{Remark 2:} For a frame to represent Euclidean geometry, it is not
required that the connection 1-forms be identically zero. It is only required
that the structure equations have a zero source term.

We now proceed to obtain the Euler-Lagrange equations or the condition that
determines the critical points of the functional $I$. We use the Gateaux
derivative for this purpose. Let $\epsilon$ denote a small parameter and
$\hat{(.)}$ the direction in which the change in the functional $I$ is
computed; this change is often referred to as the variation of
$I$.

We first calculate the variation of $I$ with respect to traction 1-forms;
$t^i\mapsto t^i+\epsilon \hat{t}^i$, where $\hat{t}^i$ are assumed to be from
the tangent space of $T^*\mathcal{S}$. The perturbed functional in the direction
of $\hat{t}^i$ can be written as,
\begin{equation}
    I(\epsilon)=\int_{\mathcal{B}}W(\theta^i) dV-((t^i+\epsilon \hat{t}^i)\otimes
    \mathsf{A}^i)\dot{\wedge}(e_j \otimes (\theta^j- \ft{d}\varphi^j))
    -\int_{\partial \mathcal{B}}\langle t^\sharp,\varphi \rangle d A.
\end{equation}
Using the definition of $\dot{\wedge}$ and Gateaux derivative, we get a vector
valued 3-form for each $i$. These three 3-forms have to be equated to zero to
get the condition for critical points in the direction of traction 1-forms.
These conditions may be formally written as,
\begin{equation}
   \begin{bmatrix}
       (\mathsf{A}^1 \wedge (\theta^1-\ed \varphi^1))
        &-(\mathsf{A}^1\wedge \ed \varphi^2)
        &-(\mathsf{A}^1\wedge\ed \varphi^2)\\
        -(\mathsf{A}^2 \wedge \ed \varphi^1)
        &(\mathsf{A}^2\wedge (\theta^2-\ed \varphi^2))
        &-(\mathsf{A}^2\wedge\ed \varphi^2)\\
        -(\mathsf{A}^3 \wedge \ed \varphi^1)
        &-(\mathsf{A}^3\wedge \ed \varphi^2)
        &(\mathsf{A}^3\wedge(\theta^3-\ed \varphi^3))
    \end{bmatrix}
    \otimes
    \begin{bmatrix}
        e_1\\
        e_2\\
        e_3
    \end{bmatrix}
    =
    \begin{bmatrix}
        0\\
        0\\
        0
    \end{bmatrix}.
\end{equation}
Since $e_i$ are orthonormal with respect to a positive definite metric, the
above equation can be true only when the coefficient matrix on the LHS is zero,
which leads to the following conditions,
\begin{subequations}
 \begin{align}
     (\mathsf{A}^1 \wedge (\theta^1-\ed \varphi^1))=0;\quad
        (\mathsf{A}^1\wedge \ed \varphi^2)=0;\quad
        (\mathsf{A}^1\wedge\ed \varphi^2)=0\\
        (\mathsf{A}^2 \wedge \ed \varphi^1)=0;\quad
        (\mathsf{A}^2\wedge (\theta^2-\ed \varphi^2))=0;\quad
        (\mathsf{A}^2\wedge\ed \varphi^2)=0\\
        (\mathsf{A}^3 \wedge \ed \varphi^1)=0;\quad
        (\mathsf{A}^3\wedge \ed \varphi^2)=0;\quad
        (\mathsf{A}^3\wedge(\theta^3-\ed \varphi^3))=0
\end{align}
\end{subequations}
Using the definition of $\mathsf{A}^i$, the above equations may be recast as,
\begin{equation}
    \begin{bmatrix}
        \theta^1-\ed \varphi^1 & \ed \varphi^2 & \ed \varphi^3\\
        \theta^1 & \theta^2-\ed \varphi^2 & \ed \varphi^3\\
        \theta^1 & \ed \theta^2 & \ed \theta^3-\varphi^3
    \end{bmatrix}
    \wedge
    \begin{bmatrix}
    \theta^2 \wedge \theta^3\\
    \theta^3 \wedge \theta^1\\
    \theta^1 \wedge \theta^2\\
    \end{bmatrix}
    =
    \begin{bmatrix}
        0\\
        0\\
        0
    \end{bmatrix}.
\end{equation}
For these equations to hold, the following
conditions must be met, 
\begin{equation}
    \theta^1-\ed \varphi^1=0;\quad
    \theta^2-\ed \varphi^2=0;\quad
    \theta^3-\ed \varphi^3=0\quad.
    \label{eq:compatabilityEL}
\end{equation} 
The above condition simply states that there exist three $0-$forms whose
exterior derivatives are the deformation 1-forms; or in other words,
the deformation 1-forms are exact and $\varphi^i$ are the potentials for the
corresponding deformation 1-forms.

We now compute the variation of $I$ with respect to the deformation $1-$forms. 
Incremental changes in the deformation 1-forms may be written as,
$\theta^i\mapsto \theta^i+\epsilon \hat{\theta}^i$, where $\epsilon
\hat{\theta}^i$ is assumed to be an element from the tangent space 
$T^*\mathcal{B}$. The perturbed functional in the direction of deformation
$1-$forms may be written as,
\begin{equation}
    I(\epsilon)=\int_{\mathcal{B}}
    W(\theta^i+\epsilon\hat{\theta}^i)d V- 
    (t^i\otimes \mathsf{A}(\epsilon)^i)\dot{\wedge}(e_j\otimes \theta^j(\epsilon))
    -\int_{\partial \mathcal{B}}\langle t^\sharp,\varphi \rangle\ed A.
\end{equation}
Using the definition of the Gateaux derivative, for each $\theta^i$ we have,
\begin{subequations}
    \label{eq:ElConstitutionRaw}
    \begin{align}
        \nonumber
        \frac{\partial W}{\partial \theta^1}
        &=[t^1(e_1){}_{\star}(\theta^2\wedge\theta^3)
        -t^2(e_1){}_{\star}(\ed \varphi^1\wedge\theta^3)
        +t^2(e_2){}_{\star}((\theta^2-\ed \varphi^2)\wedge \theta^3)
        -t^2(e_3){}_{\star}(\ed \varphi^3\wedge\theta^3)\\
        &-t^3(e_1){}_{\star}(\theta^2\wedge\ed\varphi^1)
        +t^3(e_2){}_{\star}(\theta^2\wedge\ed\varphi^3) 
        +t^3(e_3){}_{\star}(\theta^2\wedge(\theta^3-\ed \varphi^3))
        ]^\sharp\\
        \nonumber
        \frac{\partial W}{\partial \theta^2}
        &=[t^1(e_1){}_{\star}(\theta^3\wedge(\theta^1-\ed \varphi^1))
        -t^1(e_2){}_{\star}(\theta^3\wedge\ed \varphi^2)
        -t^1(e_3){}_{\star}(\theta^3\wedge\ed \varphi^3)
        -t^2(e_2){}_{\star}((\theta^3\wedge\ed \theta^1)\\
        &-t^3(e_1){}_{\star}(\ed \varphi^1\wedge\theta^1)
        -t^3(e_2){}_{\star}(\theta^2\wedge \theta^1)
        +t^3(e_3){}_{\star}((\theta^3-\ed \varphi^3)\wedge\theta^1) ]^\sharp\\
        \nonumber
        \frac{\partial W}{\partial \theta^3}
        &=[t^1(e_1){}_{\star}((\theta^1-\ed \varphi^1)\wedge\theta^2)
        -t^1(e_2){}_{\star}(\ed\varphi^2\wedge\theta^2)
        -t^1(e_3){}_{\star}(\ed\varphi^3\wedge \theta^2) 
        -t^2(e_1){}_{\star}(\theta^1\wedge\ed \varphi^1)\\
        &-t^2(e_2){}_{\star}(\theta^1\wedge(\theta^2-\ed\varphi^2))
        +t^2(e_3){}_{\star}(\theta^1\wedge\ed\varphi^3)
        +t^3(e_3){}_{\star}(\theta^1\wedge\theta^2) ]^\sharp.
    \end{align}
\end{subequations}
If we now take into account the compatibility equations previously established
in \eqref{eq:compatabilityEL}, the last equations reduce to,
\begin{subequations}
    \label{eq:ElConstitution}
    \begin{align}
        \frac{\partial W}{\partial \theta^1}
        &=[t^1(e_1){}_{\star}(\theta^2\wedge\theta^3)
        +t^2(e_1){}_{\star}( \theta^3\wedge\theta^1)
        +t^3(e_1){}_{\star}(\theta^2\wedge\theta^1)
        ]^\sharp\\
        \frac{\partial W}{\partial \theta^2}
        &=[t^1(e_2){}_{\star}(\theta^2\wedge \theta^3)
        +t^2(e_2){}_{\star}((\theta^3\wedge \theta^1)
        +t^3(e_2){}_{\star}(\theta^1\wedge \theta^2)]^\sharp\\
        \frac{\partial W}{\partial \theta^3}
        &=[t^1(e_3){}_{\star}(\theta^2\wedge \theta^3) 
        +t^2(e_3){}_{\star}(\theta^3\wedge\theta^1)
        +t^3(e_3){}_{\star}(\theta^1\wedge\theta^2) ]^\sharp.
    \end{align}    
\end{subequations}
From these equations, we see that a $2-$form accompanies the components of
traction $1-$forms; this is indeed true since we use a Piola transform to write
the constitutive rule in the reference configuration. From a comparison of
\eqref{eq:ElConstitutionRaw} and \eqref{eq:ElConstitution}, we note that the
expressions for traction in the former has additional terms. These additional
terms may be related to incompatibilities created by the emergence of defects
(such as dislocations) as the deformation evolves. In other words, without an
explicit imposition of the kinematic closure conditions (compatibility
conditions), the deformed body may never be realized as a subset of an Euclidean
space. To our understanding, this is perhaps the most significant finding of this article.

Finally, we compute the variation of $I$ with respect to deformation;
$\varphi^i\mapsto\varphi^i+\epsilon\hat{\varphi}^i$, where $\hat{\varphi}$
belongs to $T\mathcal{B}$. Using the definition of the superimposed incremental
deformation in $I$ and upon computing the Gateaux derivative, we have the
following equation,
\begin{equation}
    \label{eq:ELEquilibrioumRaw}
    \int_{\mathcal{B}}(t^i\otimes \mathsf{A}^i)\dot{\wedge}(e_j\otimes \ed\hat{\varphi}^i)=0
\end{equation}
To complete the variation, we need to shift the differential from
$\hat{\varphi}$. We first calculate the following,
\begin{equation}
    \label{eq:divTheorem}
    \ed (\varphi^k t^j(e_k) \mathsf{A}^j)=\ed \varphi^k \wedge t^j(e_k)
    \mathsf{A}^j+ \varphi^k\ed (t^j(e_k))\wedge \mathsf{A}^j+\varphi^k
    t^j(e_k)\ed\mathsf{A}^j
\end{equation}
This equation invites a few comments. The first is that we are calculating the
exterior derivative of a $2-$form, with $\varphi^k t^j(e_k)$ being a scalar. Using
the product rule of differentiation, we have expanded the right hand side of
\eqref{eq:divTheorem}. The second term in \eqref{eq:divTheorem} should be
evaluated using the connection 1-forms since it involves the exterior derivative
of a vector. This terms is relevant when one works with a frame field whose connection
1-forms are different from zero. If we invoke the compatibility of deformation, we
have $\ed \mathsf{A}^i=0$, which leaves
\eqref{eq:divTheorem} in the following form,
\begin{equation}
    \label{eq:divTheoremWithoutInc}
    \ed (\varphi^k t^j(e_k) \mathsf{A}^j)=\ed \varphi^k \wedge t^j(e_k)
    \mathsf{A}^j+ \varphi^k \ed(t^j(e_k))\wedge \mathsf{A}^j
\end{equation}
An expression similar to \eqref{eq:divTheoremWithoutInc} was utilized by Kanso
\textit{et al.} \cite{kanso2007} to define the mechanical equilibrium.  The
expression for the exterior derivative defined in \eqref{eq:divTheoremWithoutInc}
involves the connection 1-forms of the manifold, which is similar to the
covariant exterior derivatives used in gauge theories of physics \cite{hehl2012}. Using
\eqref{eq:divTheorem} in \eqref{eq:ELEquilibrioumRaw} leads to,  
\begin{equation}
    \int_{\mathcal{B}}\ed(\hat{\varphi}^k t^j(e_k)
    \mathsf{A}^j)-\hat{\varphi}^k\ed (t^j(e_k))\wedge \mathsf{A}^j=0
\end{equation}
The first term in the above equation may be converted to a boundary term via
Stokes' theorem leading to, 
\begin{equation}
    \int_{\partial\mathcal{B}}\hat{\varphi}^k t^j(e_k)
    \mathsf{A}^j-\int_{\mathcal{B}}\hat{\varphi}^k\ed (t^j(e_k))\wedge
    \mathsf{A}^j=0
\end{equation}
Using the arbitrariness of $\hat{\varphi}^k$, we conclude that,
\begin{equation}
    \ed (t^j(e_k))\wedge \mathsf{A}^j=0
\end{equation}
This is the condition for the  critical point of the mixed functional in the
direction of deformation, which represents the balance of forces.  Note that the
connection $1-$forms of the deformed configuration appear through the exterior
derivatives of the frame fields of the deformed configuration.

\subsection{Stress a Lagrange multiplier}
For a hyper-elastic solid, stress is derived from the stored energy
which may be written as a function of the deformation gradient. This assumption
permits us to write the equations of equilibrium as the Euler-Lagrange
equation of the stored energy functional. In a certain sense, the stress
generated in a hyper-elastic solid should satisfy certain integrability
condition (i.e. the existence of the stored energy function). Moreover, if
we assume the stored energy function to be translation and rotation
invariant, it implies equilibrium of forces and moments. Thus for
the hyper-elastic solid, balances of forces and moments are consequences of
translation and rotation invariance; stress is only a secondary variable
introduced for writing the equations of equilibrium in a convenient way.

When formulated as a mixed problem, the stress tensor or more specifically the
traction 1-form has a completely different role. Our mixed functional has
deformation, deformation $1-$forms and stress $2-$forms as inputs.  For the stored
energy,  viewed as a function of deformation 1-forms, translation and
rotation invariance  cannot be discussed directly, since nothing about the
geometry of the co-tangent bundle from which the deformation 1-forms were pulled
back is known. In other words, there is nothing in the stored energy function
that requires the base space for the deformed configuration to
be Euclidean. The second term in \eqref{eq:modifiedHW0} is introduced to impose
this constraint. Observe that, in \eqref{eq:modifiedHW0}, the second term is
multilinear in the input arguments, i.e., stress 2-form, differential of
deformation and deformation 1-form. The traction $1-$form may now be thought of as
a Lagrange multiplier introduced to impose the equality between the differential of
deformation and deformation 1-forms. Alternatively, the equality between the
differential of deformation and deformation 1-forms implies that the deformed
configuration is Euclidean.

\section{Discretization of differential forms}
\label{sec:discreteForms}
In this section, we consider the local finite element spaces which are suitable
for approximating differential forms over a simplicial complex. A simplicial
complex in $\mathbb{R}^m$, denoted by $K$, is a set with simplices as its
elements. By an $n$-simplex, we mean the convex hull of $n+1$ points in
$\mathbb{R}^m$. We denote a general $n$-simplex by $K^n$ and a specific
$n$-simplex by $[v_0,...,v_n]$, where $v_i\in \mathbb{R}^m$; these $v_i$ are
referred to as the vertices of the simplex. We may also call a simplex $K^n\in K$,
$n\leq m$ a $n-$face of $K$. We expect every face of $K$ to be an element in
$K$ and if two faces $K_1,K_2\in K$ intersect, then their intersection is
also a face in $K$. The dimension of $K$ is defined as the largest
dimension of the simplex contained in $K$. Finite element mesh created by
triangulating a two dimensional domain is a good example of a two dimensional
simplicial complex. Such a finite element mesh has nodes, edges and
faces, which are simplices of dimensions $0$, $1$ and $2$ respectively.
 
\begin{figure}[!h]
    \centering
    \includegraphics[scale=0.45]{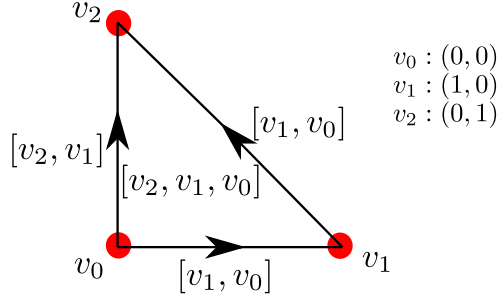}
    \caption{Reference triangle (2-simplex); a red dot indicates a vertex, while
    the arrow along an edge indicates its orientation.}
    \label{fig:referenceTriangle}
\end{figure}

From now on we restrict ourselves to two spatial dimensions; techniques
discussed here can be extended to any spatial dimensions. The just stated
assumption also permits us to work with a simplicial complex of dimension 2.
We denote the barycentric coordinates of a 2-simplex (triangle) by
$\lambda^i$, $i=0,...,2$. These coordinates satisfy the relation
$\sum_{i=0}^n\lambda^i=1$. In terms of the Cartesian coordinates, the
barycentric coordinates of the reference triangle are given by,
\begin{equation}
    \lambda^0=1-x^1-x^2;\quad\lambda^1=x^1\quad\lambda^2=x^2
\end{equation}
where, $x^1$ and $x^2$ are the Cartesian coordinates of a point within the
triangle. The reference triangle, showing the vertices and orientations of edges,
is presented in Figure \ref{fig:referenceTriangle}; we denote this reference
triangle by $\hat{K}^2$ or simply by $\hat{K}$.

\subsection{Spaces $\mathcal{P}_r\Lambda^n$ and $\mathcal{P}_r^{-}\Lambda^n$}
We denote the space of $m$ variable polynomials of degree $r$ by
$\mathcal{P}_r(\mathbb{R}^m)$. The space of polynomial differential forms
with form degree $n$ and polynomial degree $r$ is denoted by
$\mathcal{P}_r\Lambda^n(\mathbb{R}^m)$, $n\leq m$. Often we suppress
$\mathbb{R}^m$ from our notation and simply denote these spaces by
$\mathcal{P}_r$ and $\mathcal{P}_r\Lambda^n$. For vector fields,
$v_1,...,v_n \in$ $T\mathbb{R}^m$,
\begin{equation}
    \mathcal{P}_r\Lambda^n=\{\omega \in \Lambda^n(\mathbb{R}^m)| \omega(v_1,...,v_n)\in
    \mathcal{P}_r\}
\end{equation}
In other words, for a polynomial differential form $\omega$, the coefficient
functions are polynomials of degree $r$. From the definition of the spaces
$\mathcal{P}_r$ and $\mathcal{P}_r\Lambda^n$, their dimensions can be
computed as  $\binom{r+m}{m}$ and $\binom{n+r}{n}\binom{n}{k}$ 
respectively. For a differential form $\omega$ of degree $n$, the interior
product of $\omega$ with a vector $v|_{x}$, $x\in \mathbb{R}^m$, is given as,
\begin{equation}
    \kappa_{v}\omega=\omega(v,v^1,...,v^{n-1})
\end{equation}
for any vectors, $v^1,...,v^{n-1}$. From the above definition it is easy to
see that $\kappa_{v}\omega$ is a differential form of degree $n-1$. 

For any point $x\in \mathbb{R}^m$, the vector field $X\in T\mathbb{R}^m$
translates the origin to $x\in \mathbb{R}^m$. Using this vector field $X$, a
Koszul type operator on the space of polynomial differential forms with form
degree $n$ can be defined. This operator is given as,
\begin{equation}
    \kappa_X \omega=\omega(X,v^1,...,v^{n-1})
\end{equation}
where, $v^1,...,v^{n-1}$ are arbitrary vector fields form $T\mathbb{R}^m$.
From the definition, it is easy to see that $\kappa_X$ decreases the degree
of a differential form by 1 and increases the polynomial degree by 1. An
important property of $\kappa_X$ is $\kappa^2_X=0$. The operator $\kappa_X$
also commutes with affine pull-back. If $T$ is an affine linear map,
$T:\mathbb{R}^m\rightarrow \mathbb{R}^m$, then $T^{*}\kappa_X \omega =
\kappa_{X} T^{*} \omega$. The polynomial spaces $\mathcal{P}_r^-\Lambda^n$,
can be defined as,
\begin{equation}
    \mathcal{P}_r^{-}\Lambda^k=\{\omega \in \mathcal{P}_r\Lambda^k|\kappa_{X}( \omega)\in
    \mathcal{P}_{r}\Lambda^{k-1}\}
\end{equation}
The dimension of the space $\mathcal{P}_r^-\Lambda(\mathbb{R}^m)$ can be
computed as $\binom{r+k-1}{k}\binom{m+r}{m-k}$ 
which is larger than that of
$\mathcal{P}_{r-1}\Lambda^n$ but smaller than that of $\mathcal{P}_{r}\Lambda^n$. In
the case of polynomial differential forms with form degree 0, we have $\mathcal{P}_r^{-}\Lambda^0=\mathcal{P}_r\Lambda^0$;
these spaces can be identified with the Lagrange family of finite element
spaces. The spaces
$\mathcal{P}_r\Lambda^n$ and $\mathcal{P}_r^-\Lambda^n$ constitute a large
family of finite elements. Well known members of this family include the
Raviart-Thomas \cite{raviart1977} and N\'ed\'elec \cite{nedelec1986} type vector finite elements.
These subspaces of polynomial differential forms were proposed by
Arnold and co-workers \cite{arnold2006} to unify the vector finite elements used in the construction of mixed finite element techniques. These finite dimensional polynomial spaces form
the cornerstone for the finite element techniques developed under the
umbrella of finite element exterior calculus.

\subsection{Degrees of freedom and finite element bases}
In the previous subsection, we introduced the polynomial spaces
$\mathcal{P}_r^-\Lambda^n$ and $\mathcal{P}_r\Lambda^n$. We now restrict these
polynomial spaces to the reference triangle and construct basis functions
suitable for computation. The description of the computational basis functions
for the spaces $\mathcal{P}_r^-\Lambda^n$ and $\mathcal{P}_r\Lambda^n$ closely
follows the work of Arnold \textit{et al.} \cite{arnold2009geometric} where a
geometric decomposition was utilized to construct the computational basis
function on a simplex. The idea of the geometric decomposition is to index the
DoF's of a finite element (FE) space using the sub-simplices of the simplex on which the FE
space is constructed. For Lagrange finite elements over simplices, this amounts
to assigning DoF's to the vertices of the simplex. For finite elements in the
family $\mathcal{P}_r\Lambda^n$ and $\mathcal{P}_r^-\Lambda^n$, DoF's may be
indexed with edges, faces and volume. The following table gives the basis
functions for the spaces $\mathcal{P}_1\Lambda^0$, $\mathcal{P}_1\Lambda^1$ and
$\mathcal{P}_1^{-}\Lambda^1$. From Table \ref{tab:shapeFunctionsP1P1M}, it can
be seen that $\mathcal{P}_1\Lambda^0$ has DoF only on the vertex, while
$\mathcal{P}_1\Lambda^1$ has two DoF's on each edge and
$\mathcal{P}_1^-\Lambda^1$ has one DoF per edge. 
\begin{center}
    \captionof{table}{Basis functions for different FE spaces}
    \begin{tabular}{ c| c| c| c}
        \hline
        FE space & Node $[i]$ & Edge $[i,j]$ & Face $[i,j,k]$ \\
        \hline
        $\mathcal{P}_1\Lambda^0$ & $\lambda^i$ & -&-\\
        $\mathcal{P}^{-}_1\Lambda^1$ & - &$\lambda^i\ed \lambda^j-\lambda^j\ed
        \lambda^i$ &-\\
        $\mathcal{P}_1\Lambda^1$ & - &$\lambda^i\ed \lambda^j, \lambda^j\ed
        \lambda^i$ &-\\
        \hline
    \end{tabular}
    \label{tab:shapeFunctionsP1P1M}
\end{center}
\begin{figure}[!h]
    \begin{subfigure}[t]{0.33\textwidth}
        \centering
        \includegraphics[scale=0.4]{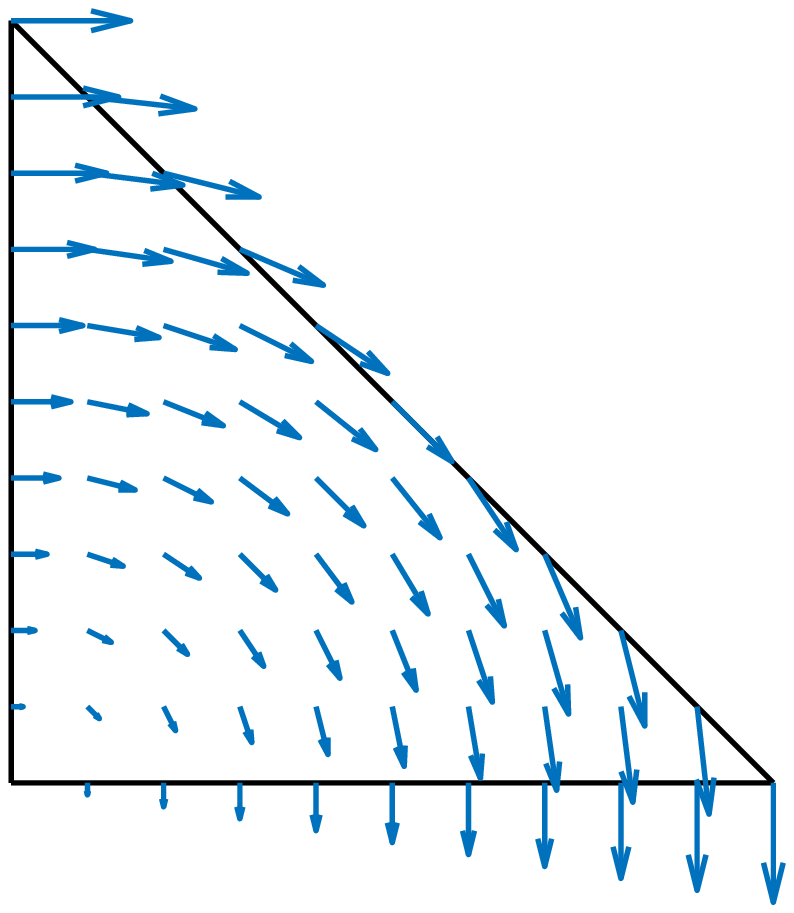}
        \caption{$\phi^1$}
    \end{subfigure}
    \begin{subfigure}[t]{0.33\textwidth}
        \centering
        \includegraphics[scale=0.4]{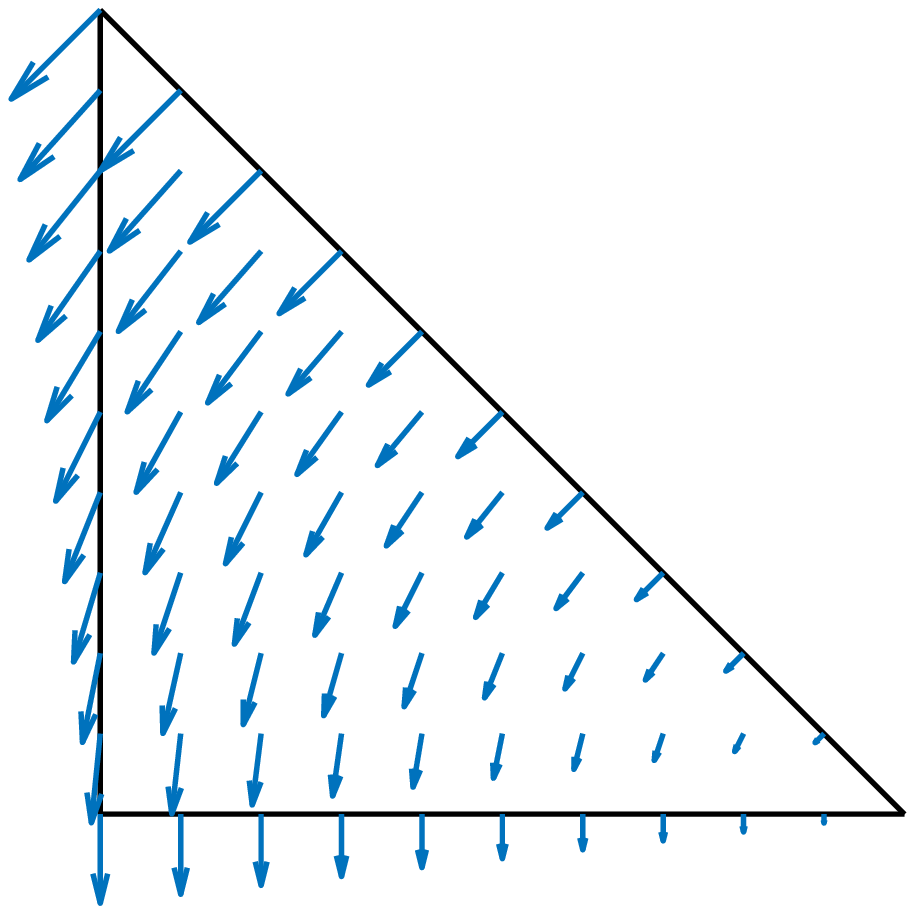}
        \caption{$\phi^2$}
    \end{subfigure}
    \begin{subfigure}[t]{0.33\textwidth}
        \centering
        \includegraphics[scale=0.4]{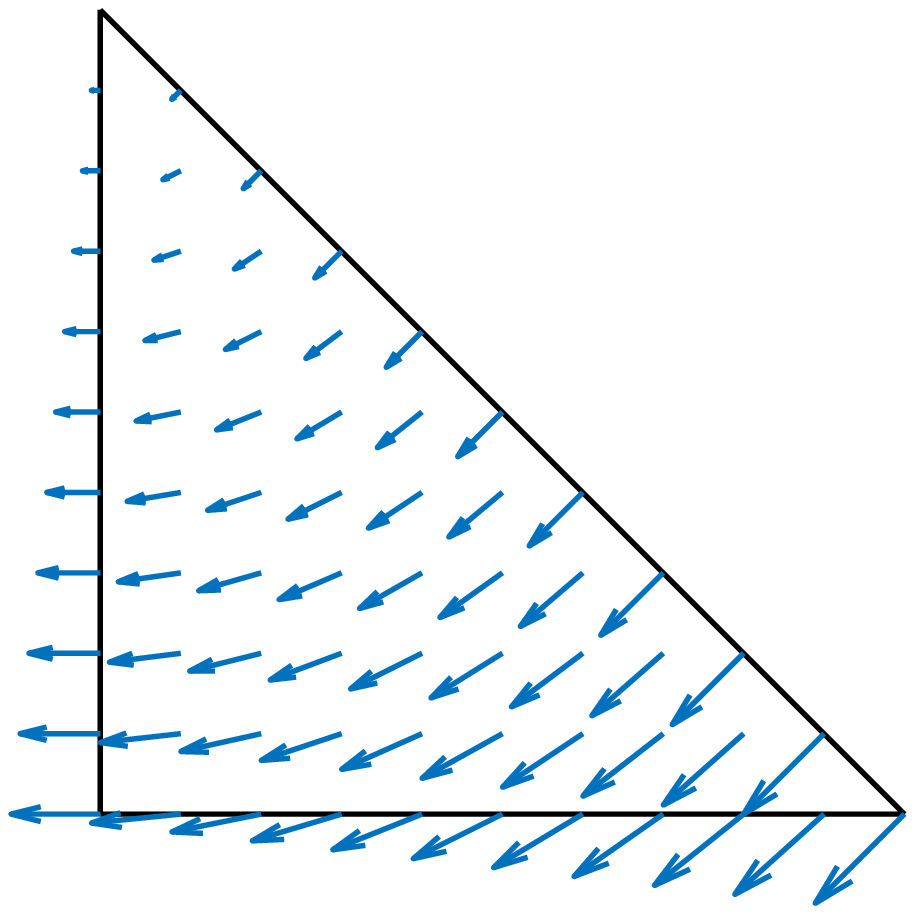}
        \caption{$\phi^3$}
    \end{subfigure}
    \caption{Vector plot of the basis functions from the space
    $\mathcal{P}_1^{-}\Lambda^1$, the 1-form basis functions are converted to
    vector fields by using the Euclidean metric.}
    \label{fig:oneFormShapeFunction}
\end{figure}

\begin{figure}[!h]
    \begin{subfigure}[t]{0.33\textwidth}
        \centering
        \includegraphics[scale=0.4]{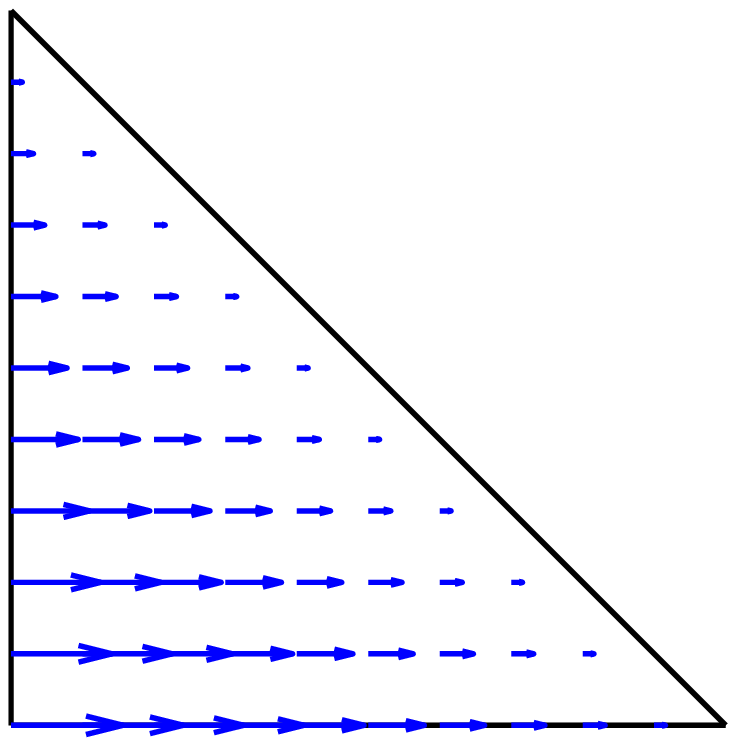}
        \caption{$\phi^1$}
    \end{subfigure}
    \begin{subfigure}[t]{0.33\textwidth}
        \centering
        \includegraphics[scale=0.4]{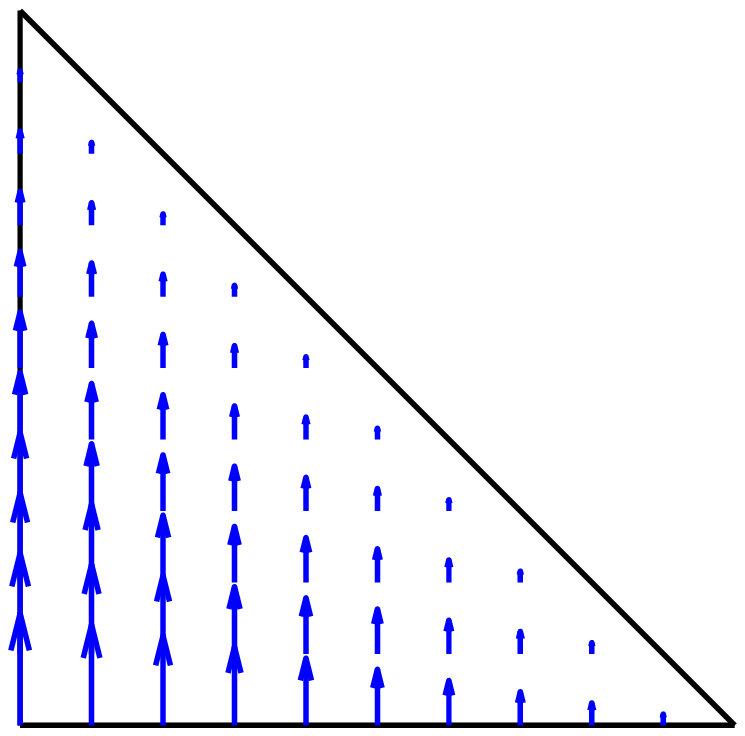}
        \caption{$\phi^2$}
    \end{subfigure}
    \begin{subfigure}[t]{0.33\textwidth}
        \centering
        \includegraphics[scale=0.4]{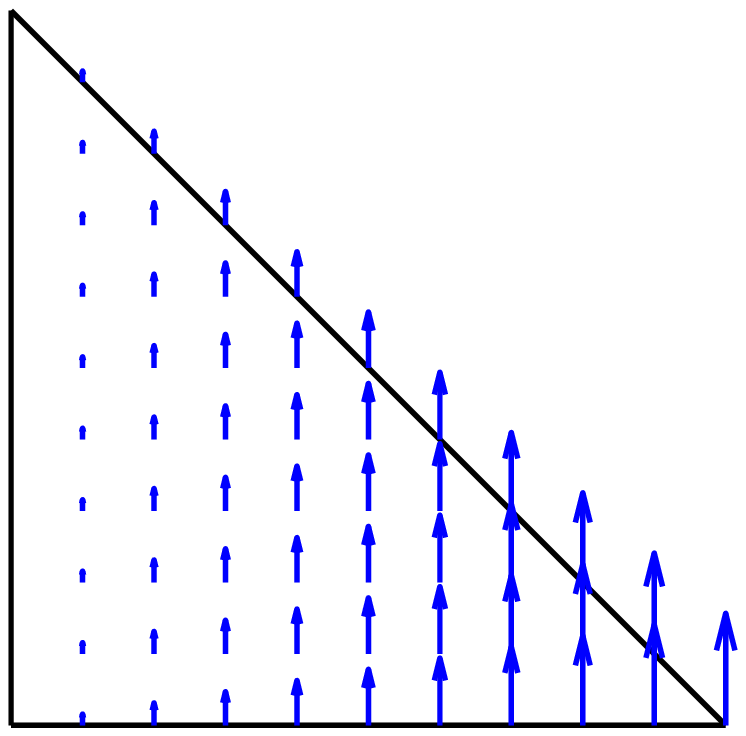}
        \caption{$\phi^3$}
    \end{subfigure}
    \begin{subfigure}[t]{0.33\textwidth}
        \centering
        \includegraphics[scale=0.4]{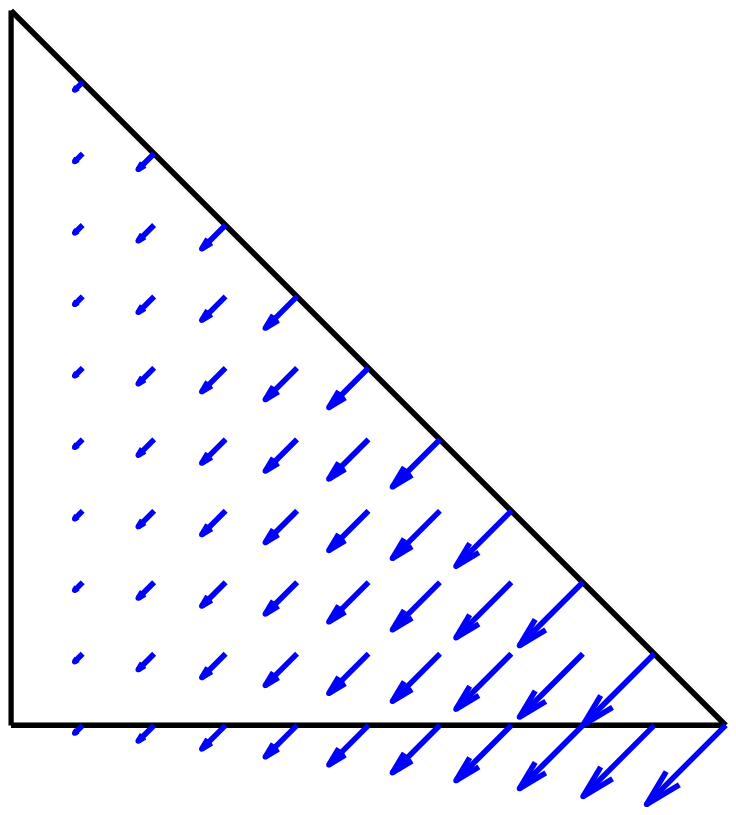}
        \caption{$\phi^4$}
    \end{subfigure}
    \begin{subfigure}[t]{0.33\textwidth}
        \centering
        \includegraphics[scale=0.4]{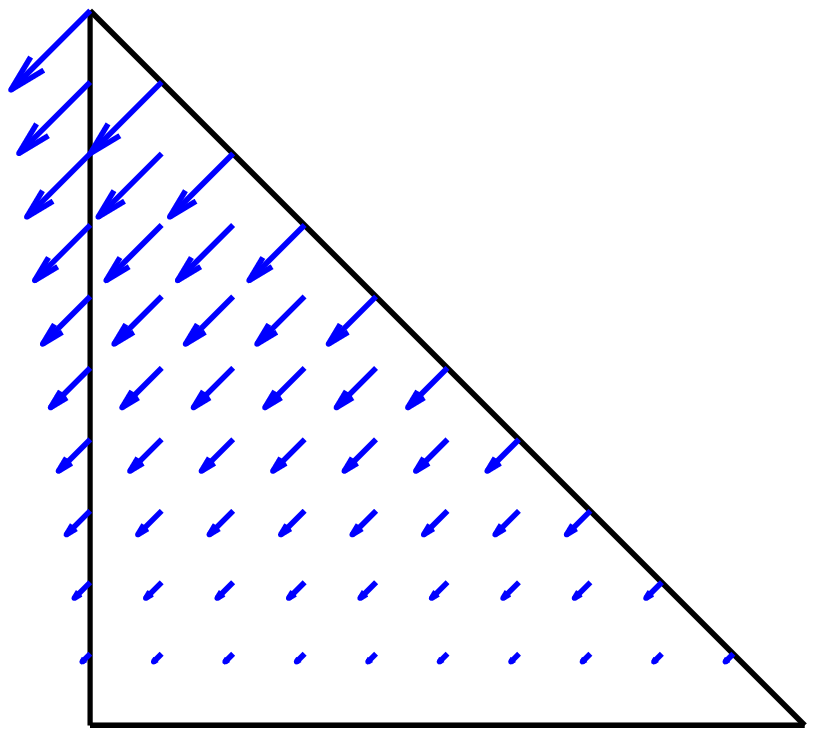}
        \caption{$\phi^5$}
    \end{subfigure}
    \begin{subfigure}[t]{0.33\textwidth}
        \centering
        \includegraphics[scale=0.4]{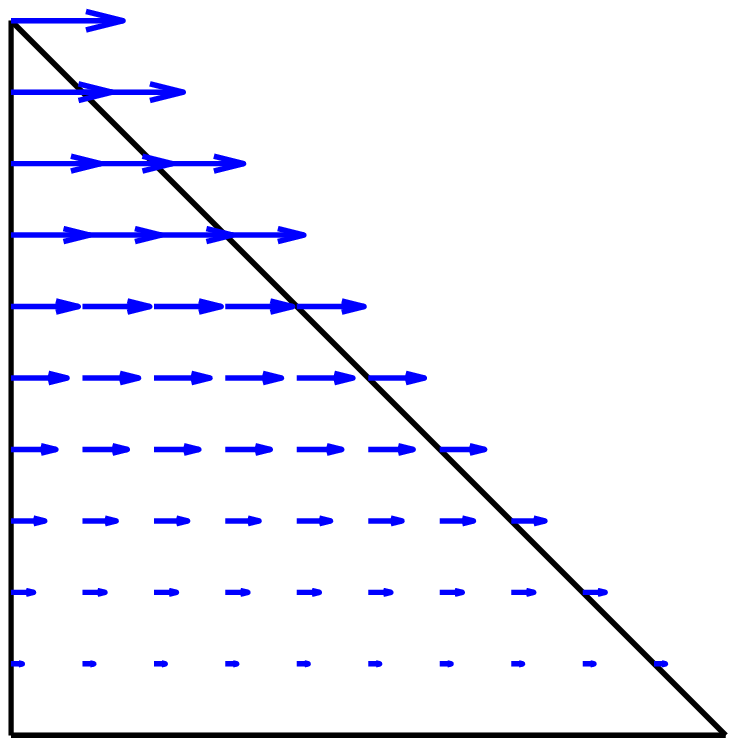}
        \caption{$\phi^6$}
    \end{subfigure}
    \caption{Vector plot of the basis functions from the space
    $\mathcal{P}_1\Lambda^1$; the 1-form basis functions are converted to
    vector fields by using the Euclidean metric.}
    \label{fig:shapeFunctionP1P1}
\end{figure}
\subsection{Whitney forms and $\mathcal{P}_1^{-}\Lambda^n$}
Differential forms of degree $n$ are functions that take $n$ vectors and produce
real numbers. Alternatively, one can also view an $n-$form as a real number
defined on a plane spanned by $n$ tangent vectors. The latter definition is more
suitable for the construction of discrete differential forms on a simplicial
complex.  On a $k-$ simplex, a discrete differential form of form degree $n$ can
be thought as a real number defined on the $k-$subsimplex. Using this idea,
Hirani \cite{hiraniThesis} constructs a discrete analog of the exterior calculus.
For such a discrete approximation, Whitney \cite{whitney2012} gave a formula for interpolating
these differential forms on a  $k-$ simplex. By producing
basis functions whose degrees of freedom are the real numbers defined on
subsimplices, Whitney was able to interpolate these differential forms within
the simplex. The basis functions, constructed by Whitney to approximate
differential forms over a simplex are now-a-days referred to as Whitney forms.  In
terms of the barycentric coordinates, these polynomial differential forms are
given by the formula,
\begin{equation}
    \phi^n_i=r!\sum_{\epsilon_i\in C(k,r)}(-1)^i \lambda_{\epsilon_i}\left(\ed \lambda_{\epsilon_0}\wedge...\wedge\widehat{\ed \lambda}_{\epsilon_i}\wedge...\wedge \ed \lambda_{\epsilon_r}\right).
    \label{eq:whitneyMap}
\end{equation}
In the above expression $C(k,r)$ is the $r$ combination of $k$-element sets from $\{1,...,k\}$.  The superscript $n$ in $\phi^n_i$ indicates the degree of the differential form and $i$ indicates the $i$th basis function.  The symbol $\widehat{(.)}$ means that the term should be removed.  Arnold showed that the space spanned by Whitney forms is exactly $\mathcal{P}_1^-\Lambda^n$ \cite{arnold2006}.

Using Whitney's forms, we now produce basis functions for the space
$\mathcal{P}_1^-\Lambda^n$ for different values $0\leq n\leq 2$. These basis
functions can be used to interpolate differential forms of degrees $0,...,2$ defined
on a 2-simplex. Whitney forms reduce to the usual linear Lagrange basis functions
on a triangle for the case $n=0$. For $n=1$, Whitney forms produce three basis functions; they
span the space $\mathcal{P}_1^-\Lambda^1$. These basis functions are given by,
\begin{equation}
    \phi_1^1=\lambda_1\ft{d}\lambda_2-\lambda_2\ft{d}\lambda_1;\quad
    \phi_2^1=\lambda_2\ft{d}\lambda_3-\lambda_3\ft{d}\lambda_2;\quad
    \phi_3^1=\lambda_3\ft{d}\lambda_1-\lambda_1\ft{d}\lambda_3
    \label{eq:basisOneForm}
\end{equation}
For differential forms of degree 2, the space of Whitney forms produces one basis function given by,
\begin{equation}
    \phi^2_1=2\left(-\lambda_1(\ft{d}\lambda_2\wedge\ft{d}\lambda_3)+\lambda_2(\ft{d}\lambda_1\wedge\ft{d}\lambda_3)-\lambda_3(\ft{d}\lambda_1\wedge\ft{d}\lambda_2)\right)
\end{equation}
Table \ref{tab:P1Dimension} gives a summary of the dimensions
and locations of the DoF's for the spaces $\mathcal{P}_1^-\Lambda^n$,
for different values of $n$ on a $2-$simplex.
\begin{center}
    \captionof{table}{Dimensions of spaces $\mathcal{P}^-_1\Lambda^n$ on a 2-simplex}
    \label{tab:P1Dimension}
    \begin{tabular}{ c| c| c }
        \hline
        Form degree & Location of DoF & Dimension of FE space \\
        \hline
        0-form                & Nodes                    & 3                              \\
        1-form                & Edges                    & 3                              \\
        2-form                & Face                     & 1                              \\
        \hline
    \end{tabular}
\end{center}

\subsection{Coordinate transformation for the FE basis}
The FE spaces we have just discussed are affine invariant
\cite{arnold2010}. This property allows us to construct the finite element basis
functions in terms of the barycentric coordinates and then use an affine transformation
to write the shape functions in terms of the Cartesian coordinates. 
The coordinate transformation between Cartesian coordinates and barycentric
coordinates of a triangle is given by,
\begin{equation}
    \begin{bmatrix}
        x^1 \\x^2
    \end{bmatrix}
    =\lambda^1 P_1+\lambda^2 P_2 + \lambda^3 P_3
    \label{eq:BarycentricToCartesian}
\end{equation}
Here $P_i$ are the vertices of the triangle and $x^i$ denote the Cartesian
coordinates of the triangle given by the vertices $[P_1,P_2,P_3]$. From the
above equation, the relationship between the differentials of the Cartesian and
barycentric coordinates can be expressed as, 
\begin{equation}
    \begin{bmatrix}
        \ft{d}x^1 \\
        \ft{d}x^2
    \end{bmatrix}
    =(P_1-P_3)\ft{d}\lambda^1 + (P_2-P_3)\ft{d}\lambda^2.
\end{equation}
In writing the above equation, we have eliminated $\lambda^3$ using $\lambda^1+\lambda^2+\lambda^3=1$. In matrix form, the above relation can be written as,
\begin{equation}
    \begin{bmatrix}
        \ft{d}x^1 \\\ft{d}x^2
    \end{bmatrix}
    =T    \begin{bmatrix}
        \ft{d}\lambda^1 \\\ft{d}\lambda^2
    \end{bmatrix},
    \label{eq:diffCoordinates}
\end{equation}
where the matrix $T$ has $P_1-P_3$ and $P_2-P_3$ as its columns. From the
constraint involving the barycentric coordinates, we have,
\begin{equation}
    \ed \lambda^1+\ed \lambda^2+\ed \lambda^3=0   
\end{equation}
In Table \ref{tab:shapeFunctionsP1P1M}, the shape functions for the $1-$form
spaces, $\mathcal{P}^1_1\lambda^1$ and $\mathcal{P}^{1-}_1\lambda^1$ are
presented in the barycentric coordinate system. These shape functions can be
transformed to the Cartesian coordinate system using 
\ref{eq:BarycentricToCartesian} and \ref{eq:diffCoordinates}. These
calculations used to transform the $1-$form basis functions are consistent with the laws of transformation of $1-$forms \cite{rognes2010}. 
\section{Discretization of modified mixed functional}
We now undertake a discrete approximation for the mixed functional discussed
in Section \ref{sec:variational}. To  achieve this, we first construct discrete approximations to the configurations and fields defined on it. The configurations of the body are discretized using a simplicial approximation, with $K(\mathcal{B})$ and $K(\mathcal{S})$ denoting the simplicial approximation for $\mathcal{B}$ and $\mathcal{S}$ respectively.  In a 2-dimensional case, a simplicial approximation amounts for placing a triangular finite element mesh on a configuration.   The objective now is to find a simplicial map, which can produce $K(\mathcal{S})$ given $K(\mathcal{B})$, material and boundary data.  By a simplicial map we mean a function which sends the vertices of $K(\mathcal{B})$ to the vertices of $K(\mathcal{S})$. In the present study, we require the simplicial map to preserve the topology of $K(\mathcal{B})$, i.e. the connectivity of the edges and faces should be preserved by the deformation.

The next step in discretizing the new functional is to get a discrete
approximation for the fields defined on the two configurations of the
body. An important point here is that, the proposed variational principle
has differential forms (not functions) as input.  In a conventional (scalar) FE
method, Lagrange basis functions  may be used to approximate the trial solution
on a $2-$simplex, DoF's are often associated with the vertices. In the
present context, we understand scalar valued functions as $0-$forms and a
conventional FE approximation (using Lagrange basis functions) as a finite
dimensional approximation of differential forms of degree zero.    As in
the conventional finite element method, constructing an FE approximation for a
differential form amounts to identifying a suitable finite dimensional
approximation space and enumerating a basis set for this space. Basis
functions and DoF's associated with the finite dimensional
approximation of differential forms defined on an $n-$simplex have been discussed in
Section \ref{sec:discreteForms}. We use these finite dimensional spaces of
differential forms to discretize our variation principle
\ref{sec:variational}. Since we are dealing with two dimensional problems, we
expect the approximation for deformation to be in a suitable finite dimensional
subspace of $\Lambda^0 \times \Lambda^0$. The approximate deformation and
traction 1-forms should be from a subspace contained in $\Lambda^1$. Yavari and
co-workers\cite{angoshtari2017,shojaei2018,shojaei2019} have presented an alternative
approach to the discretization of an HW-type variational principle; they construct
tensorial analogues of Raviart-Thomas and N\'ed\'elec finite elements to
discretize the Piola stress and deformation gradient. In contrast, the variational principle
considered in this work permits us to use FEEC directly, thus avoiding the
questionable use of tensor product finite elements.  

In the following, we use FE spaces with polynomial degree $1$ to
approximate the mixed variational principle.  A summary of different finite
element spaces used to approximate the different fields is presented in Table
\ref{tab:FEspaceDiffEle}.
\begin{center}
    \captionof{table}{FE spaces used to approximate different fields}
    \begin{tabular}{ l| c| c}
        \hline
        Field              & FE Space                     & \# DoF per Element \\
        \hline
        Displacement       & $\mathcal{P}_1\Lambda^0$     & 3                  \\
        Deformation 1-form & $\mathcal{P}_1\Lambda^1$     & 6                  \\
        Traction 1-form    & $\mathcal{P}^{-}_1\Lambda^1$ & 3                  \\
        \hline
    \end{tabular}
    \label{tab:FEspaceDiffEle}
\end{center}
Using the FE approximation for a $1-$form, the
deformation 1-forms $\theta^i$ may be written as,
\begin{equation}
    \theta_h^i=\sum_{j=1}^{n}\hat{\theta}_{j}^i\psi_j,
    \label{eq:FDApproximationForm}
\end{equation}
where, $\psi_j$ span the space $\mathcal{P}_1{\Lambda^1}$.  A subscript $h$ is
utilised in the above equation to indicate that the right hand side is only an
approximation to $\theta^i$. This finite dimensional approximation may be
conveniently written as,
\begin{equation}
    \theta_h^i=\bs{\psi}\bs\theta^i.
\end{equation}
where, $\bs \psi$ is a matrix with $\psi^i$ as its columns,
$\bs{\theta}^i$ is a vector containing the DoF's of $\theta^i$  as its
components.   Similarly, the finite dimensional approximation for the traction
1-forms may be written as,
\begin{equation}
    t^h_i=\sum_{j=1}^{n}\hat{t}^i_j\phi_j,
\end{equation}
Here, $\phi_j$ are $1-$forms spanning the space $\mathcal{P}_r^{-}\Lambda^1$ and
$\hat{t}^i_j$ are the DoF's associated with $t^i$. In matrix form,
the above approximation becomes,
\begin{equation}
    t^h_i=\bs\phi \tb t^i,
\end{equation}
where, $\bs \phi$ is a matrix with the basis of $\mathcal{P}_r^{-}\Lambda^1$ as
its columns. The finite dimensional approximation for deformation and its
differential can be written as,
\begin{equation}
    \varphi^i_h=\sum^{n}_{j=1}\hat{\varphi}^i_j\lambda^j; \quad
    \ed \varphi^i_h=\sum^{n}_{j=1}\hat{\varphi}^i_j\ed\lambda^j,
\end{equation}
where, $\lambda^j$ denotes the Lagrange basis functions and $\hat{\varphi}^i_j$
denotes the DoF associated with the $i$th deformation component.
In matrix form, the finite dimensional approximation of the differential of
deformation can be written as,
\begin{equation}
    \ed \varphi^i_h=\tb N \bs{\varphi}^i,
\end{equation}
Here, $\tb N$ denotes the matrix with exterior derivatives of the barycentric coordinates as its columns and $\bs \varphi^i$ is the DoF vector
associated with $\varphi^i$. Having introduced the discrete approximation for
configuration and fields, we may now write the discrete approximation for the mixed
functional as,
\begin{equation}
    I^h=\int_{K(\mathcal{B})}W^h(\theta^i_h)d V-(t^i_h\otimes
    d \mathsf{A}^i_h)\dot{\wedge}(E_i\otimes(\theta^i_h-\ed \phi^i_h))
    -\int_{K(\mathcal{\partial B})}\langle t^\sharp,\bs{\varphi}
    \rangle d A.
    \label{eq:HWvariationalPrinciple}
\end{equation}
In the last equation $t$ is the
traction impressed on the boundary $\partial \mathcal{B}$ of $\mathcal{B}$. Our
aim here is to construct numerical approximations for the 2-dimensional case. Recall
that for an $n-$dimensional body, stress is a covector valuded differential form
of degree $n-1$. Hence the Piola stress becomes a co-vector valued 1-form, the area-forms
in the Piola stress can be identified as $d \mathsf{A}^1=\theta^2$ and $d
    \mathsf{A^2}=\theta^1$. Incorporating these details in
\eqref{eq:HWvariationalPrinciple}, the mixed functional for the 2-dimensional case
can be written as,
\begin{equation}
    I^h=\int_{K(\mathcal{B})}W^h(\theta^1_h,\theta^2_h) dA-
    \int_{K(\mathcal{B})}(t^1_h\otimes \theta^2_h+t^2_h\otimes \theta^1_h)\dot{\wedge}
    (E_1\otimes(\theta^1_h-\ed \phi^1_h)+E_2\otimes(\theta^2_h-\ed \phi^2_h))
    -\int_{K(\mathcal{\partial B})}\langle t^\sharp,\bs{\varphi}
    \rangle dL.
    \label{eq:discreteHW2D}
\end{equation}
In the above equation, $dL$ denotes the infinitesimal line element of the
boundary curve $\partial\mathcal{B}$.

\subsection{Residue and tangent operator}
We first recall the definition of Gateaux derivative. Let
$I^h:\mathcal{V}_n\times...\times\mathcal{V}_n\rightarrow \mathbb{R}$ be a real
valued function defined on the product space
$\mathcal{V}_1\times....\times\mathcal{V}_n$, each $\mathcal{V}_i$ being a vector
space of finite or infinite dimension. If we pick elements
$(w_1,...,w_i,...,w_n)\in
    \mathcal{V}_1\times...\times\mathcal{V}_i\times...\times \mathcal{V}_n$, the
Gateaux derivative of $I$ in the direction $v_i\in \mathcal{V}_i$ is denoted by
$\mathsf{D}_{w_i}I[v_i]$ and is given by the following limit,
\begin{align}
    \nonumber
    \mathsf{D}_{w_i} I[v_i] & =\lim_{\epsilon \rightarrow 0} \frac{1}{\epsilon}[I(w_1,...,(w_i+\epsilon
        v_i),...,w_n)-I(w_1,...,w_i,...,w_n)]                                                           \\
                            & = \frac{d}{d \epsilon}[I(w_1,...,(w_i+\epsilon
        v_i),...,w_n)-I(w_1,...,w_i,...,w_n)]
    \label{eq:GateauxDerivative}
\end{align}
The derivative is thus evaluated at $(w_i,...,w_n)$ and it produces an element from
$\mathcal{V}^*_i$.  Similarly, one may also define the second derivative of $I$
by applying the above definition twice. The second derivative of $I^h$ is thus
denoted by $\mathsf{D}_{w_i}\mathsf{D}_{w_j}I^h$. We now apply the definition of Gateaux derivative to compute the first and the second
derivatives of the discrete functional.  As discussed earlier, the
discrete functional is obtained by restricting the spaces $\mathcal{V}_i$ to
a suitable finite dimensional subspace; we denote these spaces by
$\mathcal{V}_i^h$. Formally, the discrete variational functional $I^h$ can be
written as,
$I^h:\mathcal{V}_{\theta^1}^h\times\mathcal{V}_{\theta^2}^h
    \times\mathcal{V}_{t^1}^h\times\mathcal{V}_{t^2}^h\times\mathcal{V}^h_{\varphi^1}\times\mathcal{V}^h_{\varphi^2}\rightarrow
    \mathbb{R}$; here $\mathcal{V}_{\theta^i}^h$, $\mathcal{V}_{t^i}^h$ and
$\mathcal{V}^h_{\varphi^i}$ denote the finite dimensional approximation spaces
for deformation 1-form, traction 1-form and deformation respectively.
In the discrete functional, we do not discretize the frame fields $E^i$ since
we choose a fixed Cartesian frame for the reference configuration.  
The dimension of the product space  on which the extremization problem is posed
largely depends on the FE mesh and the finite dimensional approximation space
utilised to discretize different fields. Strain
energy functions associated with non-linear elasticity is non-quadratic; thus
finding extremisers for the discrete functional now falls within the realm of
numerical optimization. We utilise Newton's method to numerically find the
extremizers of the discrete functional. Newton's method involves the
computation of the first and second derivatives associated with the discrete
functional given in \ref{eq:discreteHW2D}. In the finite element literature, the
first and second derivatives of the discrete energy functional are often called
the residual vector and tangent operator respectively. The components of a
residual vector are the generalised forces acting on the respective DoF's. Thus the
global residual vector is obtained by stacking the first derivatives of the
discrete functional with respect to different DOF's one above
the other.  Formally, the residue vector can be written as,
\begin{equation}
    \mathcal{R}=\left[\mathsf{D}_{\bs \theta^1} I^h, \; \mathsf{D}_{\bs \theta^2}
        I^h, \; \mathsf{D}_{\tb t^1} I^h, \; \mathsf{D}_{\tb t^2} I^h, \; \mathsf{D}_{\bs
            \varphi^1} I^h, \; \mathsf{D}_{\bs \varphi^2} I^h\right]^t
\end{equation}
The matrix form of the tangent operator may now be written as,
\begin{equation}
    \mathcal{K}=
    \begin{bmatrix}
        \mathsf{D}_{\bs \theta^1}\mathsf{D}_{\bs \theta^1}I^h   &
        \mathsf{D}_{\bs \theta^1}\mathsf{D}_{\bs \theta^2}I^h   &
        \mathsf{D}_{\bs \theta^1}\mathsf{D}_{\tb t^1}I^h        &
        \mathsf{D}_{\bs \theta^1}\mathsf{D}_{\tb t^2}I^h        &
        \mathsf{D}_{\bs \theta^1}\mathsf{D}_{\bs \varphi^1}I^h  &
        \mathsf{D}_{\bs \theta^1}\mathsf{D}_{\bs \varphi^2}I^h    \\
                                                                &
        \mathsf{D}_{\bs \theta^2}\mathsf{D}_{\bs \theta^2}I^h   &
        \mathsf{D}_{\bs \theta^2}\mathsf{D}_{\bs t^1}I^h        &
        \mathsf{D}_{\bs \theta^2}\mathsf{D}_{\bs t^2}I^h        &
        \mathsf{D}_{\bs \theta^2}\mathsf{D}_{\bs \varphi}I^h    &
        \mathsf{D}_{\bs \theta^2}\mathsf{D}_{\bs \varphi^2}I^h    \\
                                                                &
                                                                &
        \mathsf{D}_{\tb t^1}\mathsf{D}_{\tb t^1}I^h             &
        \mathsf{D}_{\tb t^1}\mathsf{D}_{\tb t^2}I^h             &
        \mathsf{D}_{\tb t^1}\mathsf{D}_{\bs \varphi^1}I^h       &
        \mathsf{D}_{\tb t^1}\mathsf{D}_{\bs \varphi^2}I^h         \\
                                                                &
        \text{Symmetric}                                        &
                                                                &
        \mathsf{D}_{\tb t^2}\mathsf{D}_{\tb t^2}I^h             &
        \mathsf{D}_{\tb t^2}\mathsf{D}_{\bs \varphi^1}I^h       &
        \mathsf{D}_{\tb t^1}\mathsf{D}_{\bs \varphi^2}I^h         \\
                                                                &
                                                                &
                                                                &
                                                                &
        \mathsf{D}_{\bs \varphi^1}\mathsf{D}_{\bs \varphi^1}I^h &
        \mathsf{D}_{\bs \varphi^1}\mathsf{D}_{\bs \varphi^2}I^h   \\
                                                                &
                                                                &
                                                                &
                                                                &
                                                                &
        \mathsf{D}_{\bs \varphi^2}\mathsf{D}_{\bs \varphi^2}I^h   \\
    \end{bmatrix}
\end{equation}
It should be noted that the tangent operator is symmetric since it is obtained
as the Hessian of an energy functional. We now present the expressions for the first and second derivatives of the discrete functional.
\begin{align}
    \mathsf{D}_{\bs\theta^1}I^h   & =\int_{\mathcal{T}_\mathcal{B}}\text{D}_{\bs \theta^1}W^h
    -t^1_h(E_2)\bs{\psi}\wedge\theta^2_h+t^2_h(E_1)\bs{\psi}\wedge\theta^2_h+t^1_h(E_1)\bs{\psi}\wedge
    \ed \varphi^1_h+t^1_h(E_2)\bs{\psi}\wedge\ed \varphi^2_h                                  \\
    \mathsf{D}_{\bs \theta^2}I^h  & =\int_{\mathcal{T}_\mathcal{B}}\text{D}_{\bs \theta^2}W^h
    +t^1_h(E_2)\bs{\psi}\wedge\theta^1_h-t^2_h(E_1)\bs{\psi}\wedge\theta^1_h+t^2_h(E_1)\bs{\psi}\wedge\ed
    \phi^1_h
    +t^2_h(E_2)\bs{\psi}\wedge\ed \varphi^2_h                                               \\
    \mathsf{D}_{\tb t^1}I^h       &
    =\int_{\mathcal{T}_\mathcal{B}}\theta^1_h\wedge(\ed \varphi^2_h-\theta^2_h)\bs{\phi}^tE_2+
    (\theta^1_h\wedge\ed \varphi^1_h)\bs{\phi}^tE_1                                           \\
    \mathsf{D}_{\tb t^2}I^h       &
    =\int_{\mathcal{T}_\mathcal{B}}\theta^2_h\wedge(\ed
    \varphi^1_h-\theta^1_h)\bs{\phi}^tE_1+(\theta^2_h\wedge\ed
    \varphi^2_h)\bs{\phi}^tE_2                                                              \\
    \mathsf{D}_{\bs \varphi^1}I^h & =\int_{\mathcal{T}_\mathcal{B}}
    -t^1_h(E_1)\tb{N}\wedge\theta^1_h-t^2_h(E_1) \tb{N}\wedge\theta^2_h                       \\
    \mathsf{D}_{\bs \varphi^2}I^h & =\int_{\mathcal{T}_\mathcal{B}}
    -t^1_h(E_2)\tb{N}\wedge\theta^1_h-t^2_h(E_2) \tb{N}\wedge\theta^2_h
\end{align}
The second derivative of the discrete functional may be computed as,
\begin{align}
    \mathsf{D}_{\bs \theta^1}\mathsf{D}_{\bs \theta^1}I^h   & =\int_{\mathcal{T}_\mathcal{B}}
    \mathsf{D}_{\bs \theta^1}\mathsf{D}_{\bs \theta^1}W^h;
    \quad
    \mathsf{D}_{\bs \theta^2}\mathsf{D}_{\bs \theta^2}I^h=\int_{\mathcal{T}_\mathcal{B}}
    \mathsf{D}_{\bs \theta^2}\mathsf{D}_{\bs \theta^2}W^h                                           \\
    \mathsf{D}_{\bs \theta^2}\text{D}_{\bs \theta^1}I^h     & =\int_{\mathcal{T}_\mathcal{B}}
    \mathsf{D}_{\bs \theta^2}\text{D}_{\bs \theta^1}W^h
    +(t^2_h(E_1)-t^1_h(E_2))(\bs{\phi}\wedge\bs{\phi})                                          \\
    \mathsf{D}_{\tb t^1}\mathsf{D}_{\tb t^1}I^h             & =0;\quad
    \mathsf{D}_{\tb t^2}\mathsf{D}_{\tb t^2}I^h=0;\quad
    \mathsf{D}_{\tb t^1}\mathsf{D}_{\tb t^2}I^h=0                                             \\
    \mathsf{D}_{\bs \varphi^1}\mathsf{D}_{\bs \varphi^1}I^h & =0;\quad
    \mathsf{D}_{\bs \varphi^2}\mathsf{D}_{\bs \varphi^2}I^h=0;\quad
    \mathsf{D}_{\bs \varphi^1}\mathsf{D}_{\bs \varphi^2}I^h=0                                 \\
    \mathsf{D}_{\tb t^1}\mathsf{D}_{\bs \theta^1}I^h        & =
    \int_{\mathcal{T}_\mathcal{B}}-(\bs{\psi}\wedge\theta^2_h)\otimes\bs{\phi}^t
    E_2+(\bs{\psi}\wedge \ed \varphi^1_h)\otimes\bs{\phi}^t E_1
    +(\bs{\psi}\wedge\ed \varphi^2_h)\otimes\bs{\phi}^tE_2                                    \\
    \mathsf{D}_{\tb t^2}\mathsf{D}_{\bs \theta^1}I^h        & =\int_{\mathcal{T}_\mathcal{B}}
    (\bs{\psi}\wedge\theta^2_h)\otimes \bs{\phi}E_1;\quad
    \mathsf{D}_{\tb t^1}\mathsf{D}_{\bs \theta^2}I^h=\int_{\mathcal{T}_\mathcal{B}}
    (\bs{\psi}\wedge\theta^1_h)\otimes \bs{\psi}^tE_2                                         \\
    \mathsf{D}_{\tb t^2}\mathsf{D}_{\bs \theta^2}I^h       & =\int_{\mathcal{T}_\mathcal{B}}
    -(\bs{\psi}\wedge \theta^1_h)\otimes\bs{\phi}^tE_1+(\bs{\psi}\wedge\ed
    \varphi^1_h)\otimes \bs{\phi}^t E_1+(\bs{\psi}\wedge\ed \varphi^2_h)\otimes
    \bs{\phi}^tE_2                                                                            \\
    \mathsf{D}_{\bs \varphi^1}\mathsf{D}_{\bs \theta^1}I^h  & =
    \int_{\mathcal{T}_\mathcal{B}}t^1_h(E_1)(\bs{\psi}\wedge \bs{N});\quad
    \mathsf{D}_{\bs \varphi^2}\mathsf{D}_{\bs \theta^1}I^h=
    \int_{\mathcal{T}_\mathcal{B}}t^1_h(E_2)(\bs{\psi}\wedge\bs{N})                           \\
    \mathsf{D}_{\bs \varphi^2}\mathsf{D}_{\bs \theta^1}I^h  & =
    \int_{\mathcal{T}_\mathcal{B}}t^2_h(E_1)(\bs{\psi}\wedge\bs{N});\quad
    \mathsf{D}_{\bs \varphi^2}\mathsf{D}_{\bs \theta^2}I^h=
    \int_{\mathcal{T}_\mathcal{B}}t^2_h(E_2)(\bs{\psi}\wedge\bs{N})                           \\
    \mathsf{D}_{\bs \varphi^1}\mathsf{D}_{\bs{t}^1}I^h      & =
    \int_{\mathcal{T}_\mathcal{B}}-\bs{\phi}E_1\otimes(\bs{N}\wedge\theta^1_h);\quad
    \mathsf{D}_{\bs \varphi^2}\mathsf{D}_{\bs{t}^1}I^h=
    \int_{\mathcal{T}_\mathcal{B}}-\bs{\phi}E_2\otimes(\bs{N}\wedge\theta^1_h)                \\
    \mathsf{D}_{\bs \varphi^1}\mathsf{D}_{\bs{t}^2}I^h      & =
    \int_{\mathcal{T}_\mathcal{B}}-\bs{\phi}E_1\otimes(\bs{N}\wedge\theta^2_h);\quad
    \mathsf{D}_{\bs \varphi^2}\mathsf{D}_{\bs{t}^2}I^h=
    \int_{\mathcal{T}_\mathcal{B}}-\bs{\phi}E_2\otimes(\bs{N}\wedge\theta^2_h)
\end{align}
\section{Numerical results}
\label{sec:numericalImplementation}
We apply the mixed FE approximation based on our variational principle  to numerically study solutions of a few benchmark problems; the objective is to demonstrate its efficacy against numerical instabilities such as volume and bending locking. We consider only two dimensional problems. The material response in the simulations are calculated using a neo-Hookian type stored energy function with the density given by \cite{reese2002},
\begin{equation}
    W(\theta^1,\theta^2)=\frac{\mu}{2}(I_1-2)-\mu\ln J +\frac{\kappa}{2}(\ln J)^2.
    \label{Eq:neoHookianEnergy}
\end{equation}
A discrete approximation to the stored energy density is obtained by replacing the deformation 1-forms by their finite dimensional approximations. Using the discrete stored energy functional, the contribution of the stored energy density to the residue and tangent may be computed as the first and second derivatives of \eqref{Eq:neoHookianEnergy}. These derivatives are computed using the definition of Gateaux derivative in \eqref{eq:GateauxDerivative}. The first derivative of the stored energy function can be evaluated as,
\begin{equation}
    \mathsf{D}_{\bs{\theta}^i}W^h=\frac{\mu}{2} \mathsf{D}_{\bs{\theta}^i}I_1^h+\left(\frac{\kappa\ln J^h}{J}-\frac{\mu}{J^h}\right)\mathsf{D}_{\bs{\theta}^i}J^h.
    \label{eq:neoRes}
\end{equation}
Here, $I_1^h$ and $J^h$ denote the finite dimensional approximations for $I_1$ and $J$. The first derivative of $I^h_1$ with respect to $\bs{\theta}^i$ may be computed as,
\begin{equation}
    \mathsf{D}_{\bs{\theta}^i}I_1^h=
    2\bs \psi \bs\theta^i.
\end{equation}
Similarly, the first derivative of $J^h$ with respect to $\bs{\theta}^i$ may be computes as,
\begin{equation}
    \mathsf{D}_{\bs{\theta}^1}J^h= 
    \bs \psi \wedge \theta^2;\quad     \mathsf{D}_{\bs{\theta}^2}J^h= 
    -\bs \psi \wedge \theta^1
\end{equation}

The stiffness associated with 1-form degrees of freedom may be evaluated as,
\begin{equation}
    \mathsf{D}_{\bs{\theta}^j}\mathsf{D}_{\bs{\theta}^i}W^h=\frac{\mu}{2}\mathsf{D}_{\bs{\theta}^j}\mathsf{D}_{\bs{\theta}^i}I_1^h+
    \left[
        \frac{\mu}{(J^h)^2}+\frac{\kappa}{(J^h)^2}-\frac{\lambda \ln J^h}{(J^h)^2}
    \right]\mathsf{D}_{\bs{\theta}^j}J^h\otimes \mathsf{D}_{\bs{\theta}^i}J^h
    +\left[
        \frac{-\mu}{J^h}+\frac{\kappa \ln J^h}{J^h}
    \right]\mathsf{D}_{\bs{\theta}^j}\mathsf{D}_{\bs{\theta}^i}J^h.
    \label{eq:neoHess}
\end{equation}
Explicit expressions for the second derivatives of $I^h_1$ may be computed as,
\begin{equation}
    \mathsf{D}_{\bs{\theta}^i}\mathsf{D}_{\bs{\theta}^j}I^h_1=
    \begin{cases}
        2\bs \psi^t\bs \psi; \quad i=j \\
        0; \quad i\neq j 
\end{cases}.
\end{equation}
Similarly, the second derivatives of $J^h$ may be computed as,
\begin{equation}
    \mathsf{D}_{\bs{\theta}^i}\mathsf{D}_{\bs{\theta}^j}J^h=
    \begin{cases}
        0; \quad i=j\\
        \bs \psi \wedge \bs \psi; \quad i\neq j 
\end{cases}.
\end{equation}
The convergence of deformation 1-forms is assessed through the integral $\int_\mathcal{B}\norm{\theta^1}^2+\norm{\theta^2}^2d V$. Here $\norm{.}$ is given by the metric tensor of the reference configuration. The convergence of stress is evaluated using $\int_{\mathcal{B}}\norm{P}dV$, where $P$ is the first Piola stress in the conventional sense, which can constructed using the traction and deformation 1-forms in the following manner,
\begin{equation}
    P=t^1\otimes \theta^2+t^2\otimes \theta^1.
\end{equation}
In writing the above equation, use has been made of the definition of stress as a covector valued 1-form.

\subsection{Cook's membrane problem}
\begin{figure}[!h]
    \centering
    \begin{subfigure}[b]{0.49\textwidth}
        \includegraphics[scale=0.45]{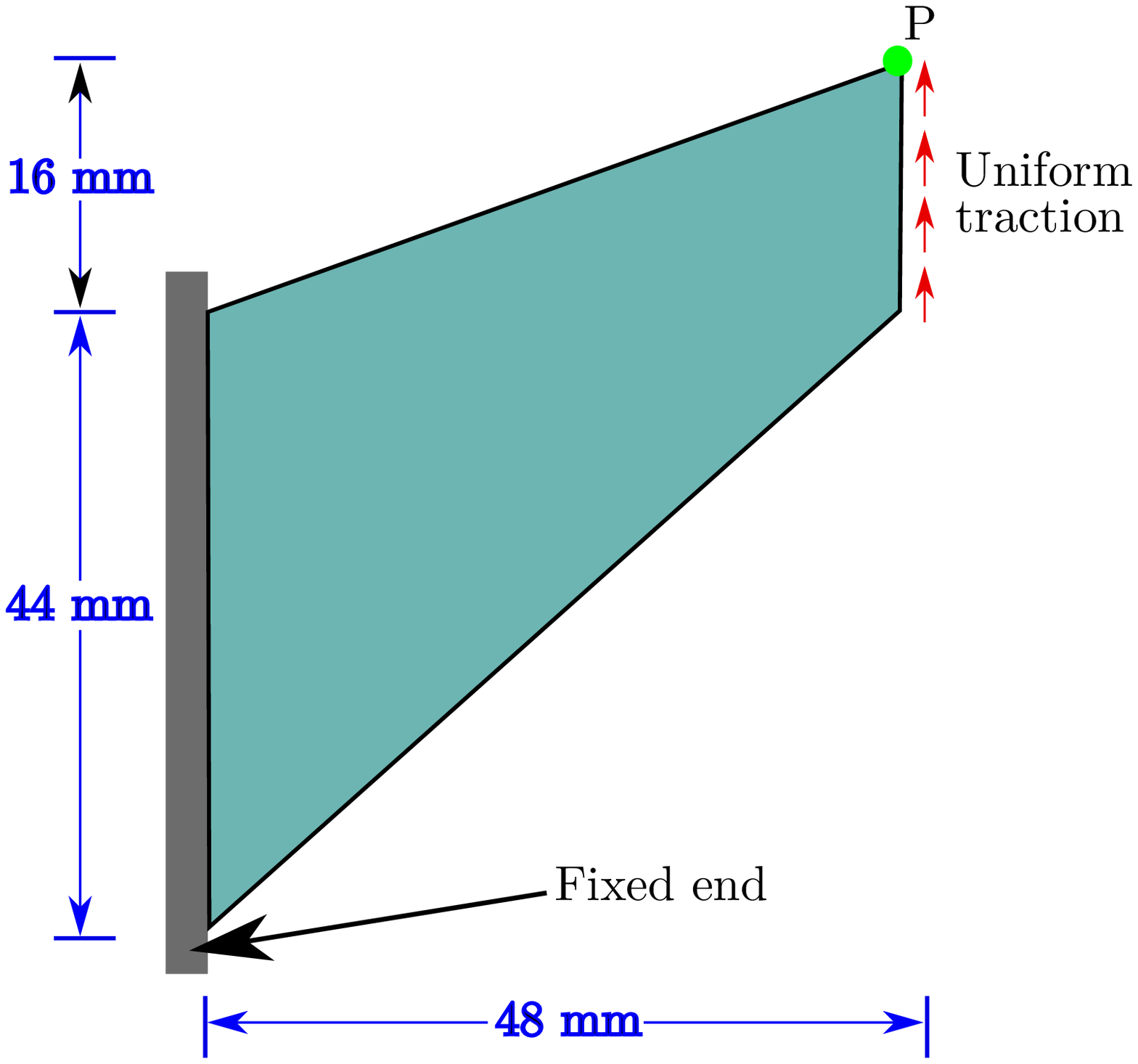}
        \caption{Cook's membrane: Dimensions of the domain along with boundary conditions;
        displacement is measured at the point P.}
         \label{fig:cookBvP}
    \end{subfigure}
    \begin{subfigure}[b]{0.49\textwidth}
        \includegraphics[scale=0.5]{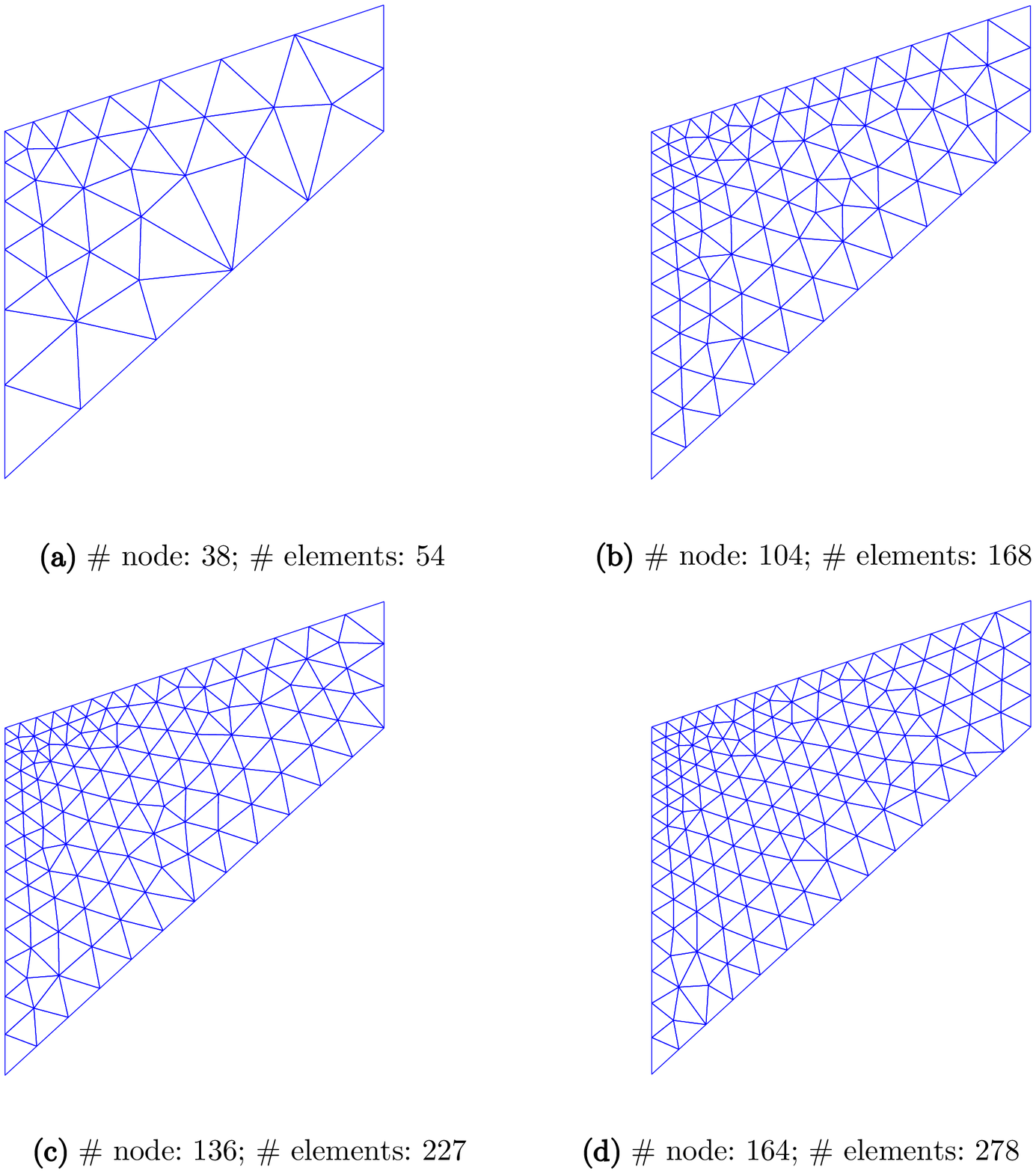}
        \caption{FE mesh used to study convergence}
        \label{fig:cookFeMesh}
    \end{subfigure}
    \caption{(a) Boundary conditions used to study the Cook's membrane; (b) representative sequence of mesh used in the convergence study}
\end{figure}

\begin{figure}[!h]
    \centering
    \begin{subfigure}[b]{0.32\textwidth}
        \includegraphics[scale=0.4]{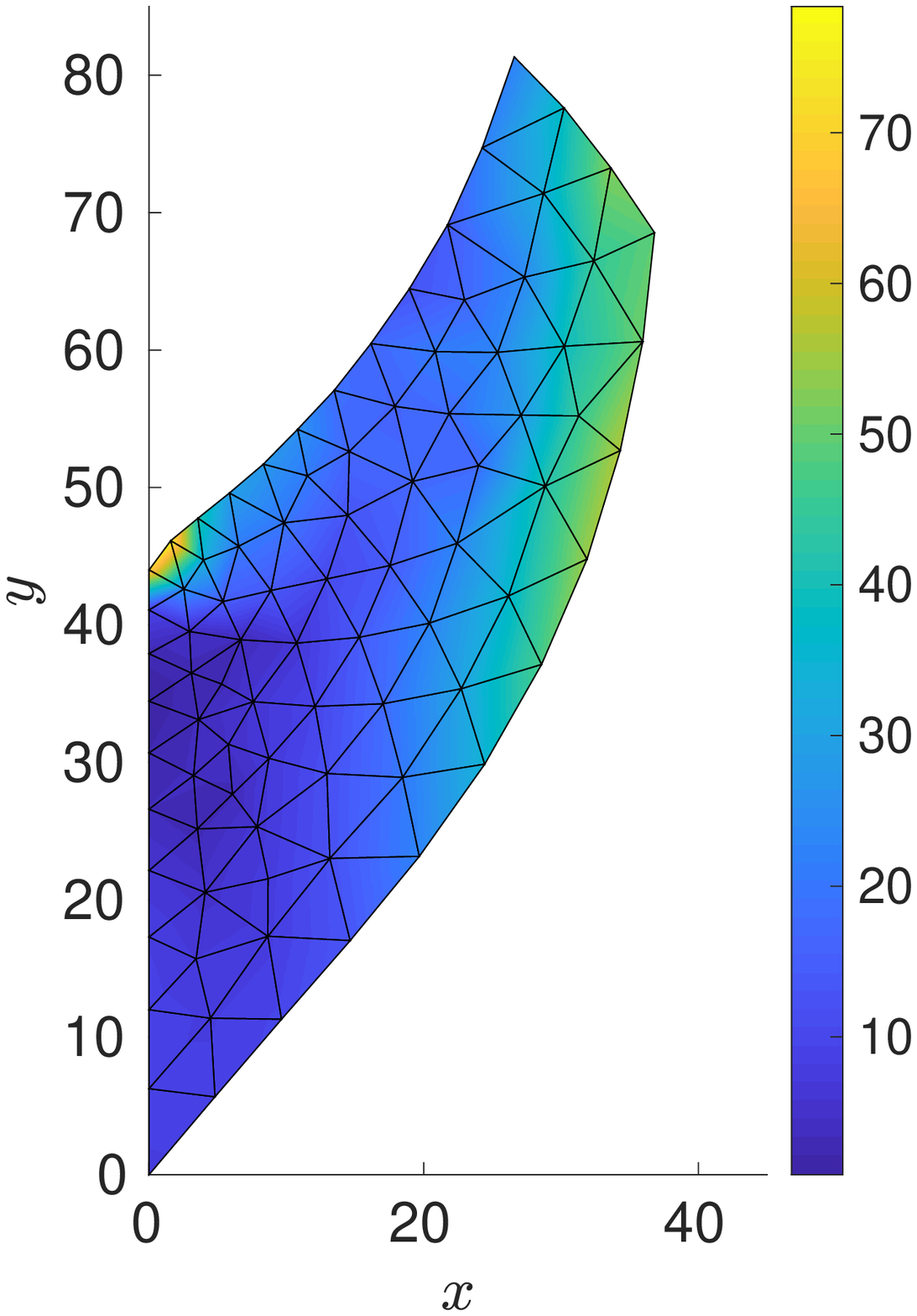}
    \end{subfigure}
    \begin{subfigure}[b]{0.32\textwidth}
        \includegraphics[scale=0.4]{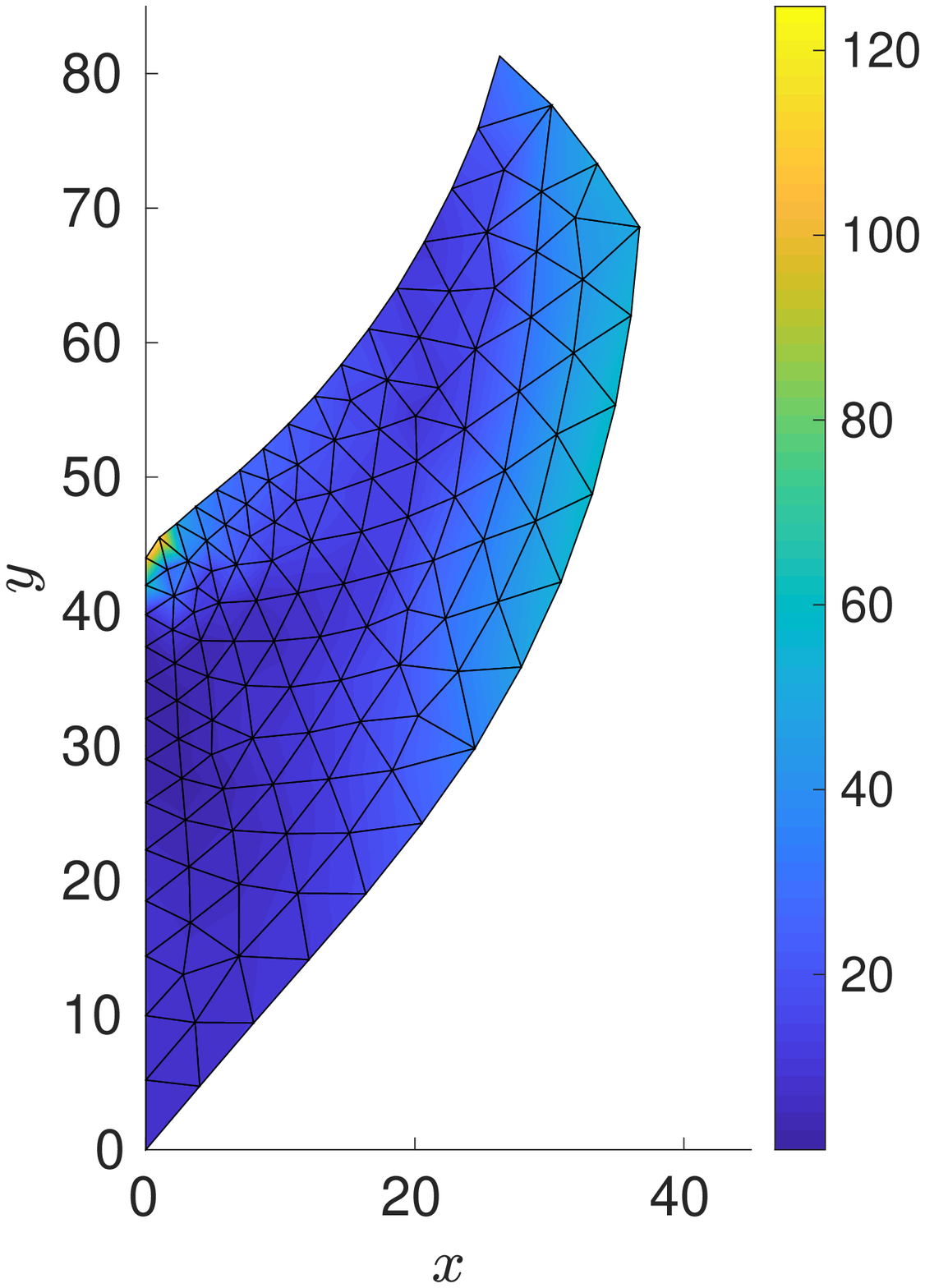}
    \end{subfigure}
    \begin{subfigure}[b]{0.32\textwidth}
        \includegraphics[scale=0.4]{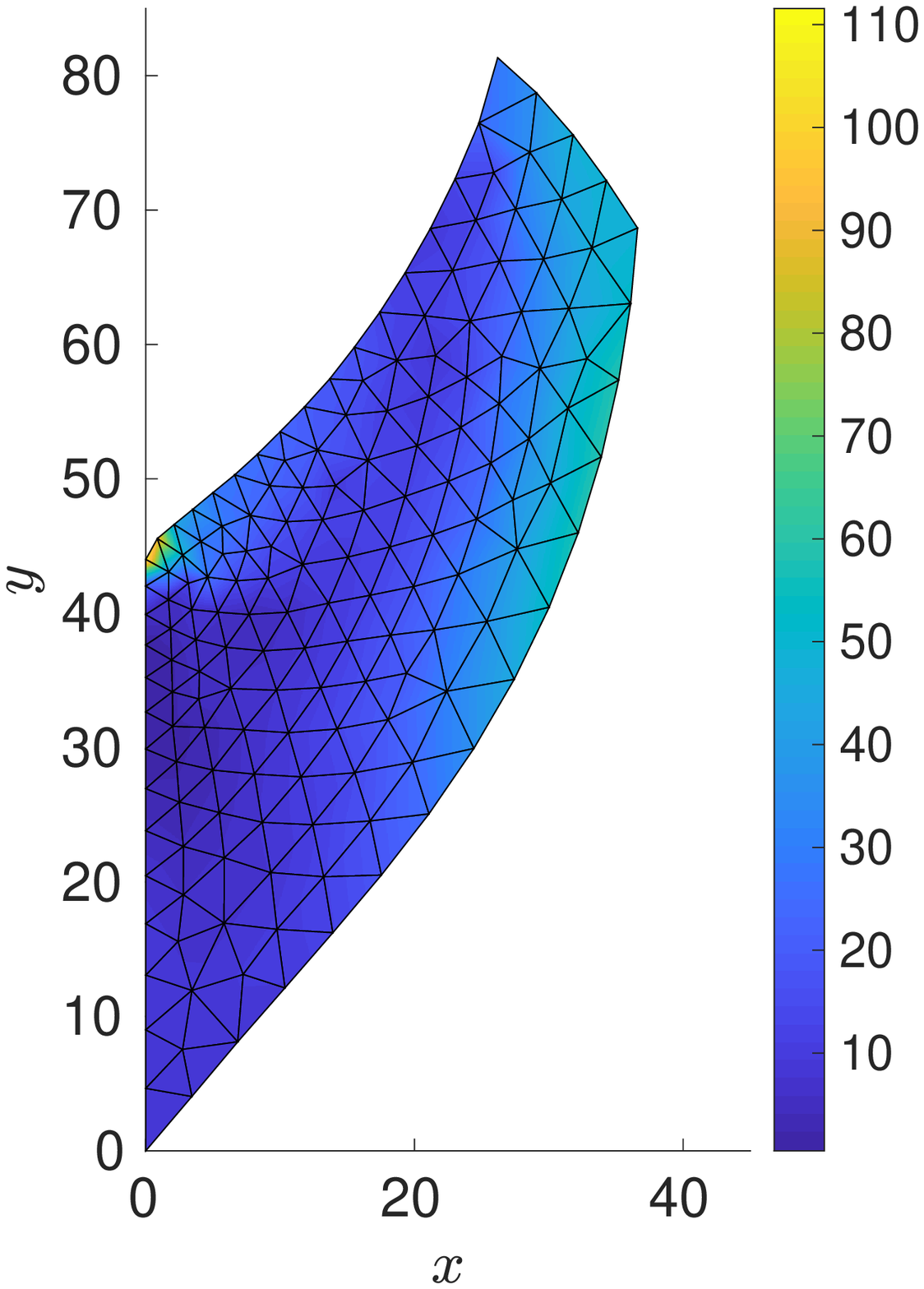}
    \end{subfigure}
    \caption{Deformed configuration of Cook's membrane for a load level of 32 $N/mm^2$. The colors indicate the norm of the first Piola stress.}
    \label{fig:cookStressContour}
\end{figure}

\begin{figure}
    \centering
    \includegraphics[scale=0.5]{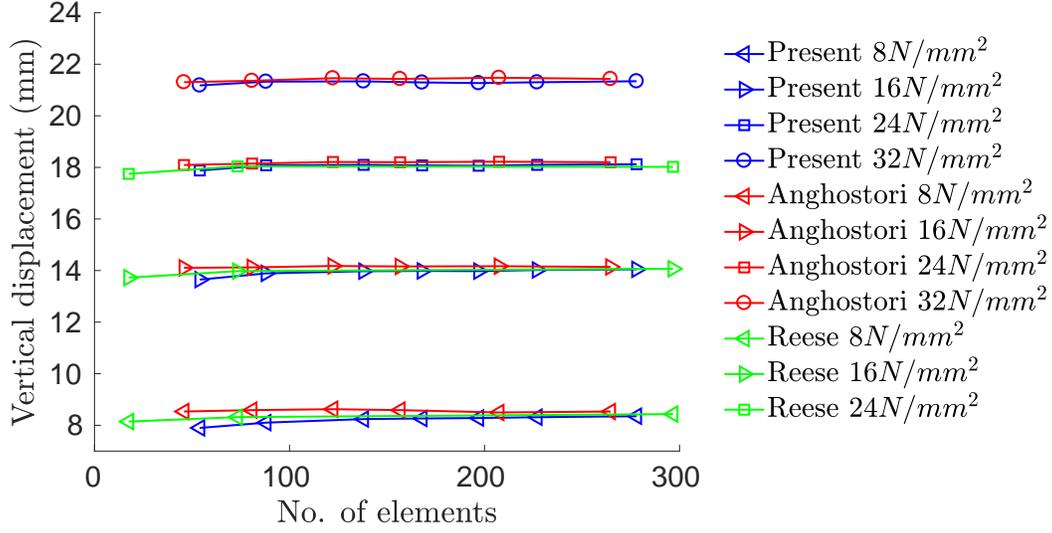}
    \caption{Convergence of displacement at point $P$ for different load
        levels and mesh refinements. For comparison, we have plotted the
        convergence curves given in Reese and Wriggers
        \cite{reese2000} (in green) and Angoshtari \textit{et al.} \cite{angoshtari2017}(in red)}
        \label{fig:cookTipConvergence}
\end{figure}

\begin{figure}
    \begin{subfigure}[b]{0.49\textwidth}
        \includegraphics[width=\textwidth]{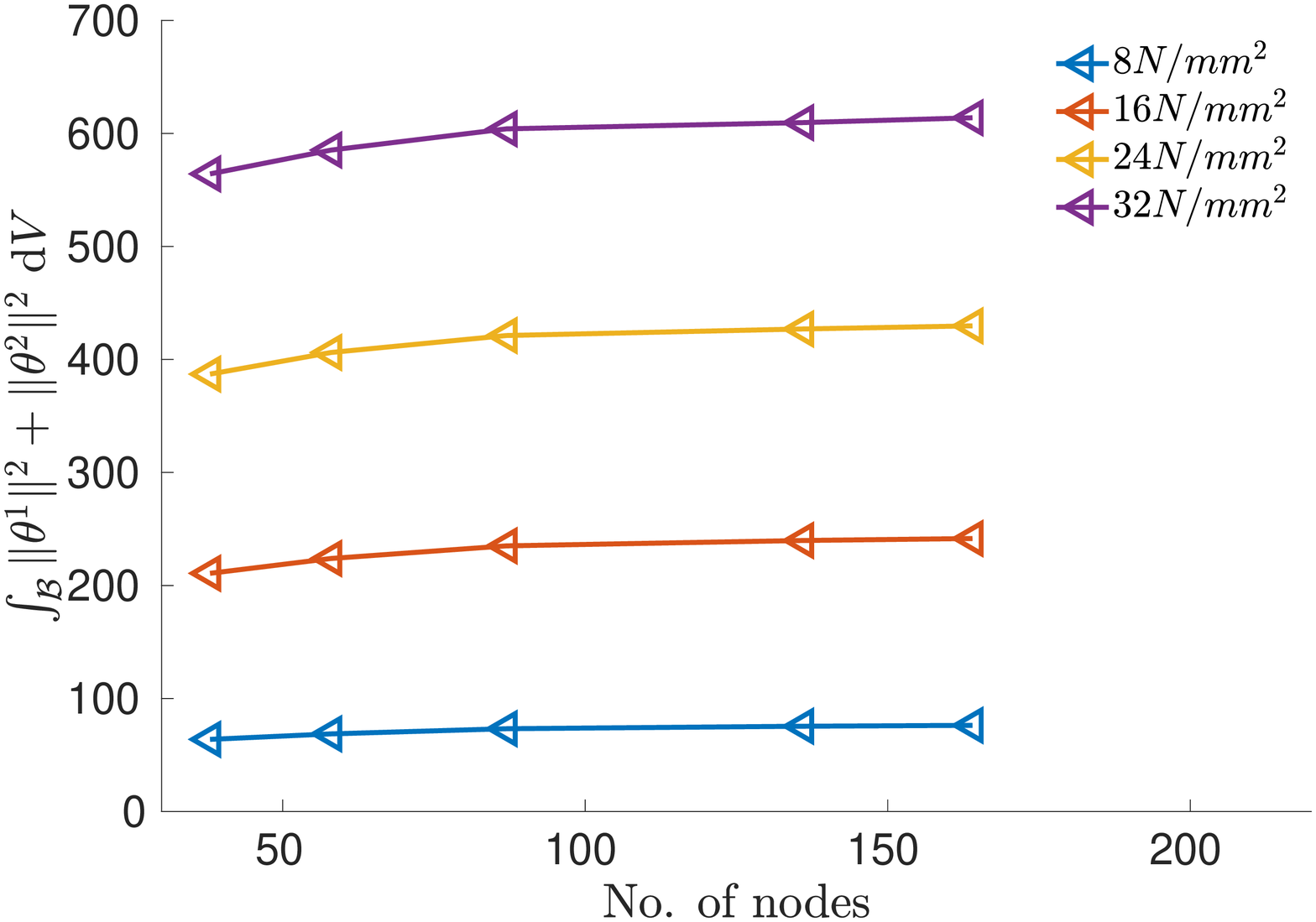}
        \caption{Convergence of deformation 1-forms}
    \end{subfigure}
    \begin{subfigure}[b]{0.49\textwidth}
        \includegraphics[width=\textwidth]{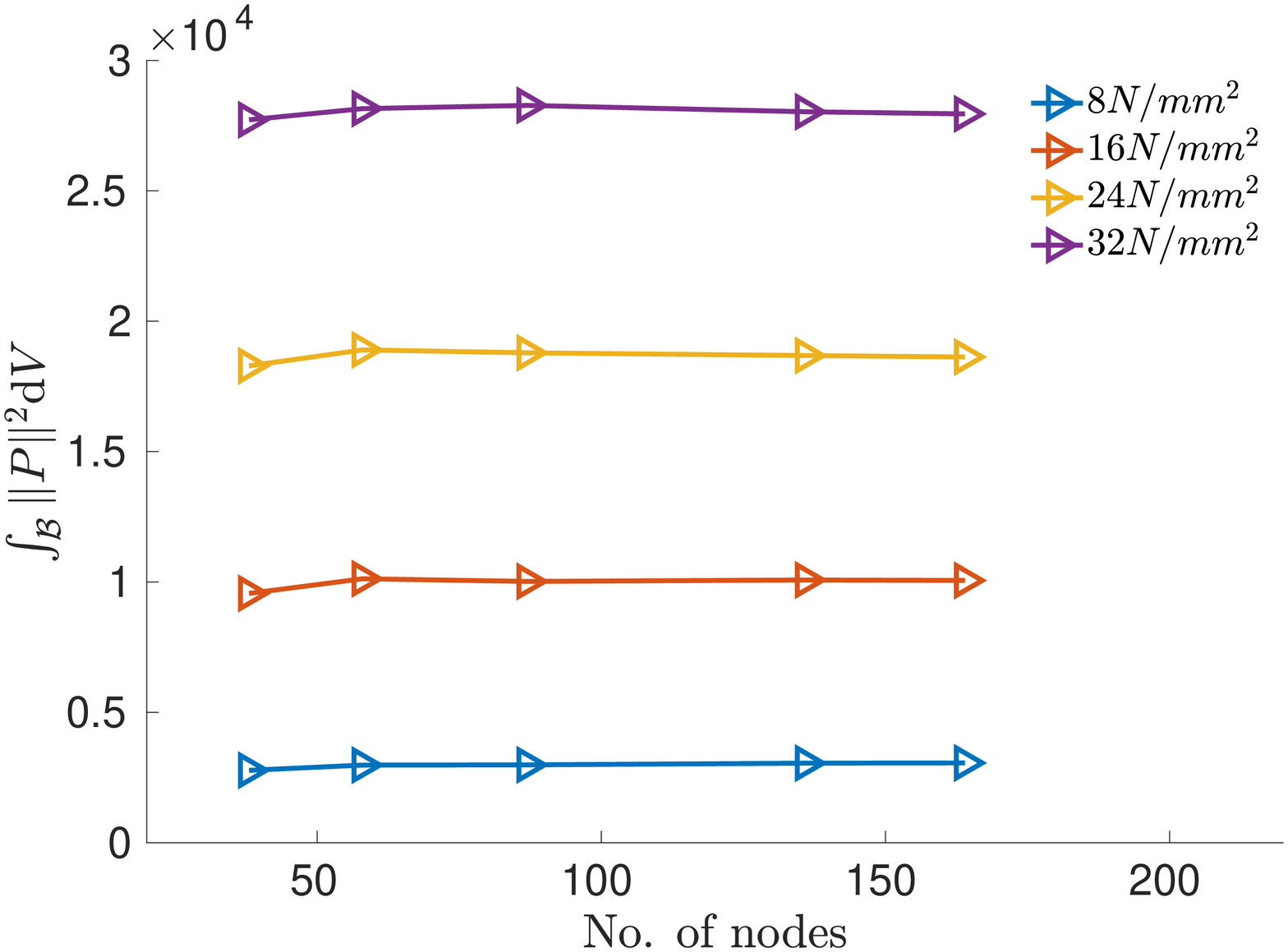}
        \caption{Convergence of first-Piola stress}
    \end{subfigure}
    \caption{Convergence of deformation 1-form and first-Piola stress for different load levels}
    \label{fig:cookThetaStressConver}
\end{figure}

We first study the performance of our FE formulation against bending locking at the incompressible limit. Cook's membrane is a standard benchmark problem used to test the efficiency of an FE scheme against bending induced numerical instabilities. Cook's membrane is a trapizoidal cantilever beam; the domain and boundary conditions used in the simulations are shown in Fig. \ref{fig:cookBvP}. A representative sequence of the FE mesh used to numerically study the convergence is shown in Fig. \ref{fig:cookFeMesh}.  We choose the material constants as  $\mu=80.194$ $N/mm^2$, $\kappa=400889.8$ $N/mm^2$; these parameters correspond to a quasi-incompressible material. The convergence of the tip displacement (at point P) is shown in Figure \ref{fig:cookTipConvergence}. The convergence plot obtained by Reese \cite{reese2002} and Angoshtari \textit{et al.} \cite{angoshtari2017} are also plotted alongside. Note that we use finite elements with polynomial degree 1 in our simulations; contrast this with Angoshtari \textit{et al.} who use finite elements with polynomial degree 2 to obtain the convergence plots given in \ref{fig:cookTipConvergence}. Moreover, coefficients in the stored energy function used  by Angoshtari \textit{et al.} is different from the present work and that of Reese \cite{reese2002}; indeed, the comparison given in \cite{angoshtari2017} does not seem appropriate. Results on convergence of the deformation 1-form and first Piola stress are given in Figure \ref{fig:cookThetaStressConver}. In Figure \ref{fig:cookStressContour}, the deformed configuration predicted by our mixed FE method along with the norm of the first Piola stress is presented.


\subsection{Compression of a rectangular block}
\begin{figure}[!h]
    \centering
    \begin{subfigure}[b]{0.47\textwidth}
        \includegraphics[scale=0.5]{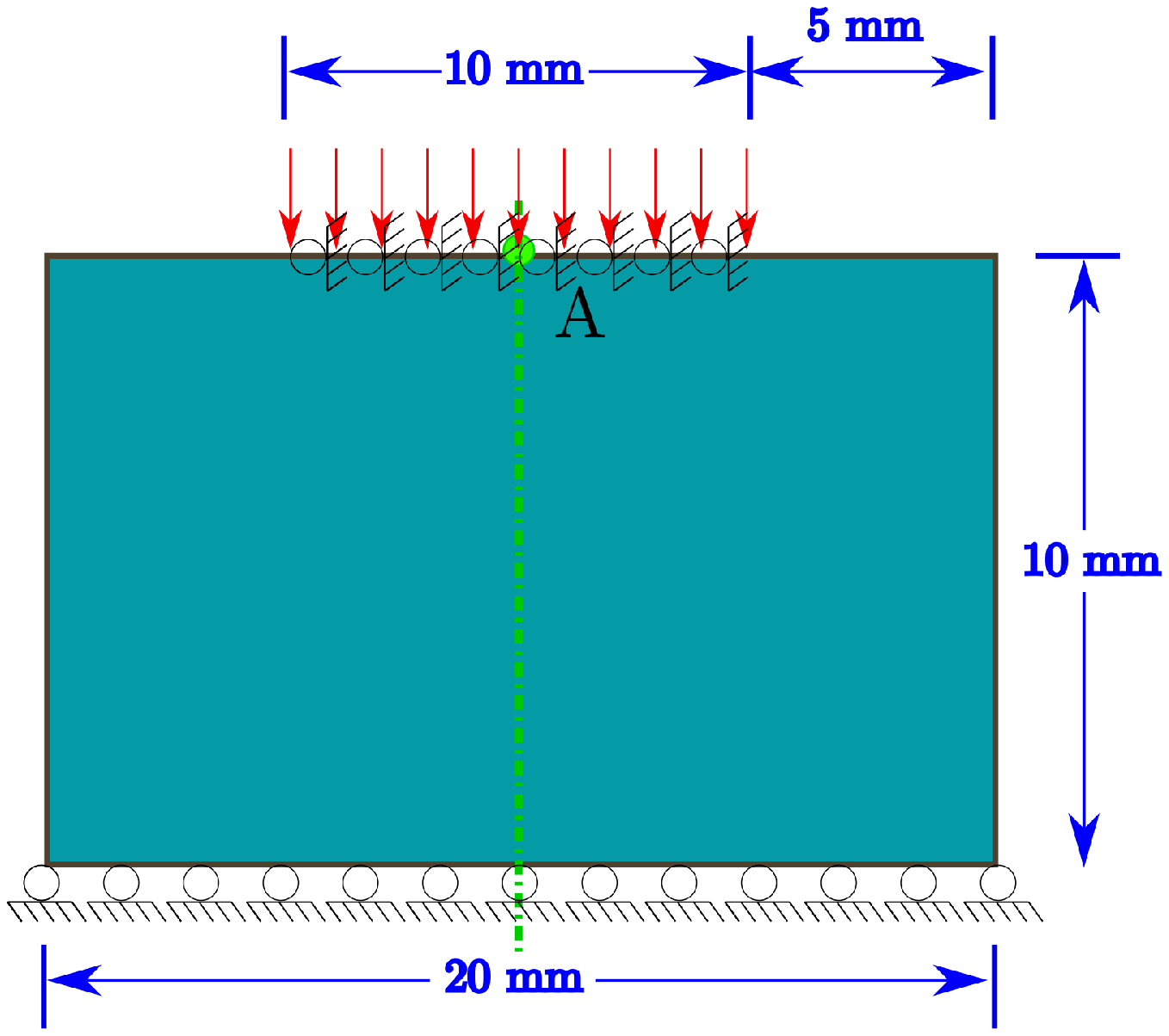}
        \caption{Boundary conditions}
    \end{subfigure}
    \begin{subfigure}[b]{0.47\textwidth}
        \includegraphics[scale=0.5]{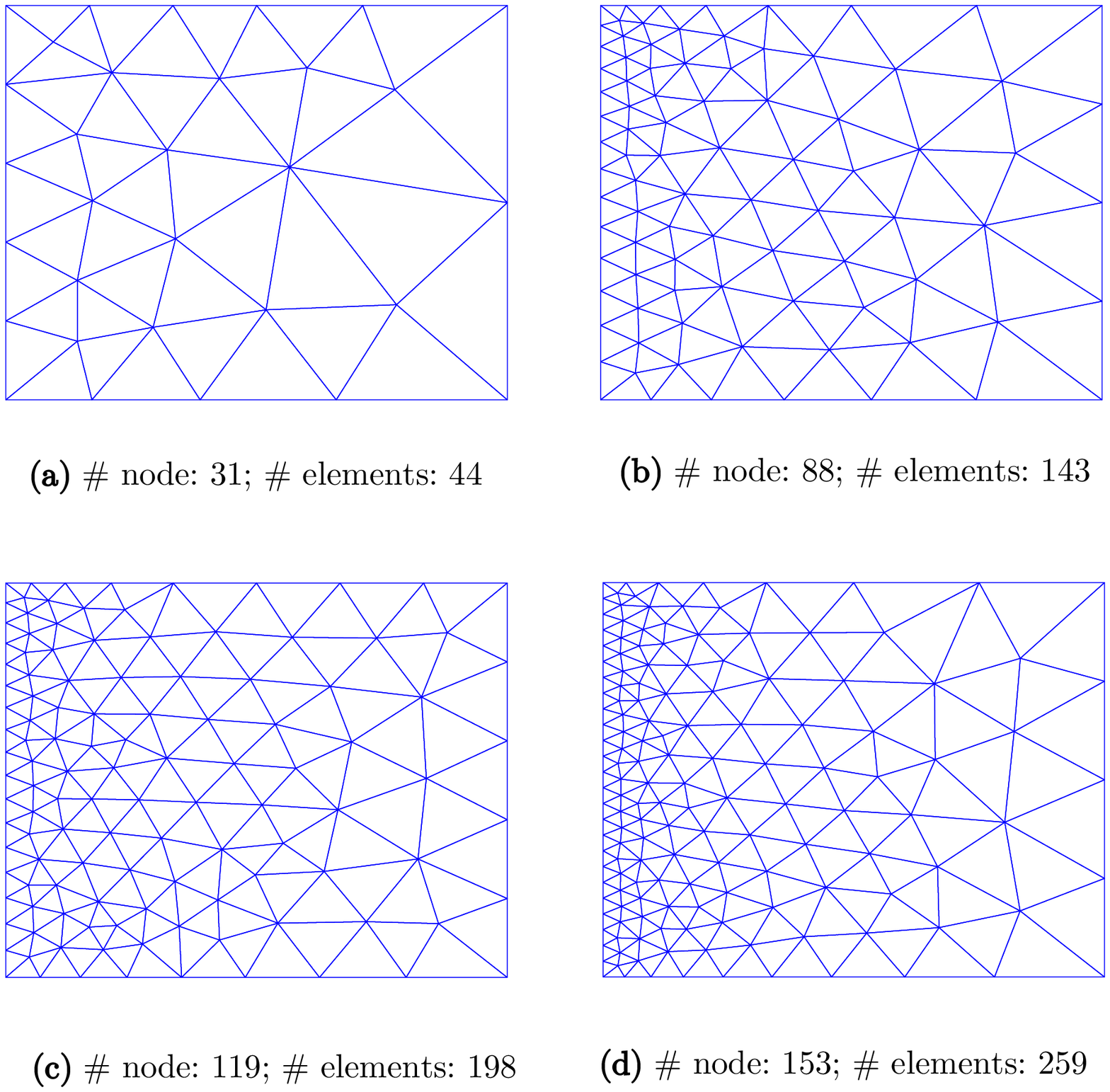}
        \caption{FE mesh used in the simulation}
    \end{subfigure}
    \caption{(a) Boundary conditions used to study the compression of the rectangular block; (b) representative sequence of mesh used in the convergence study}
    \label{fig:sqCompBvP}
\end{figure}

\begin{figure}
        \centering
        \includegraphics[scale=0.5]{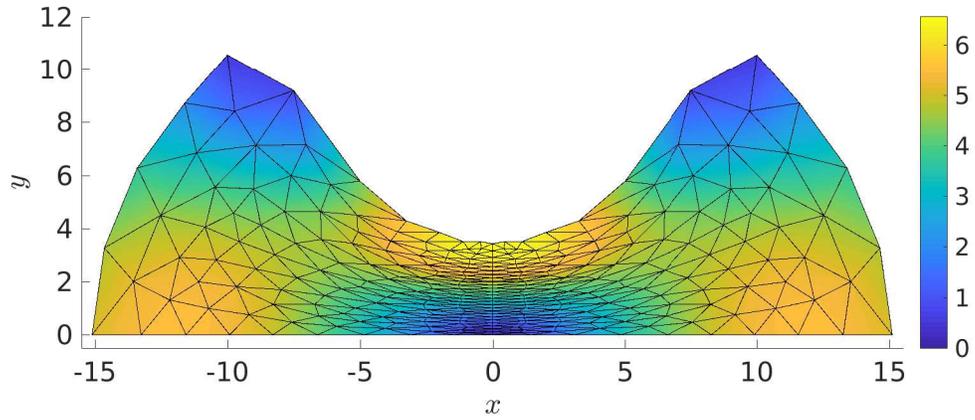}
        \caption{Deformed shape predicted by our FE simulation; the deformation is mirrored about the $y-$ axis for clarity. Colors on the deformed configuration indicate the norm of displacement.}
        \label{fig:DeformedCofigMirror}
\end{figure}

\begin{figure}
    \centering
    \includegraphics[scale=0.5]{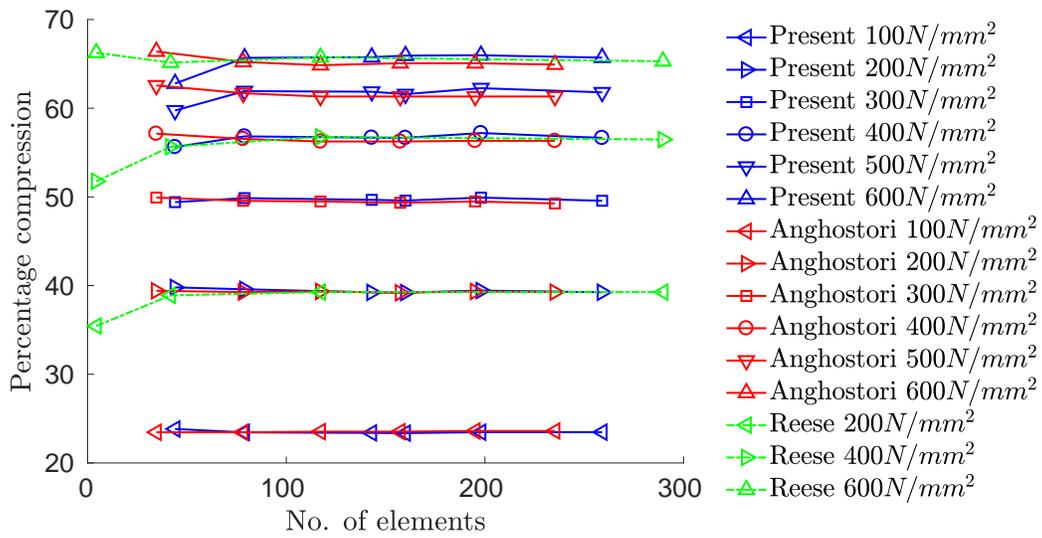}
    \caption{Convergence of displacement at the point $P$ for different load levels.  For comparison we have plotted the
    convergence results of Reese and Wriggers \cite{reese2000}  and Angoshtari \textit{et al.} \cite{angoshtari2017}.}
    \label{fig:sqCompDispConvergence}
\end{figure}

\begin{figure}
    \begin{subfigure}[b]{0.49\textwidth}
        \includegraphics[width=\textwidth]{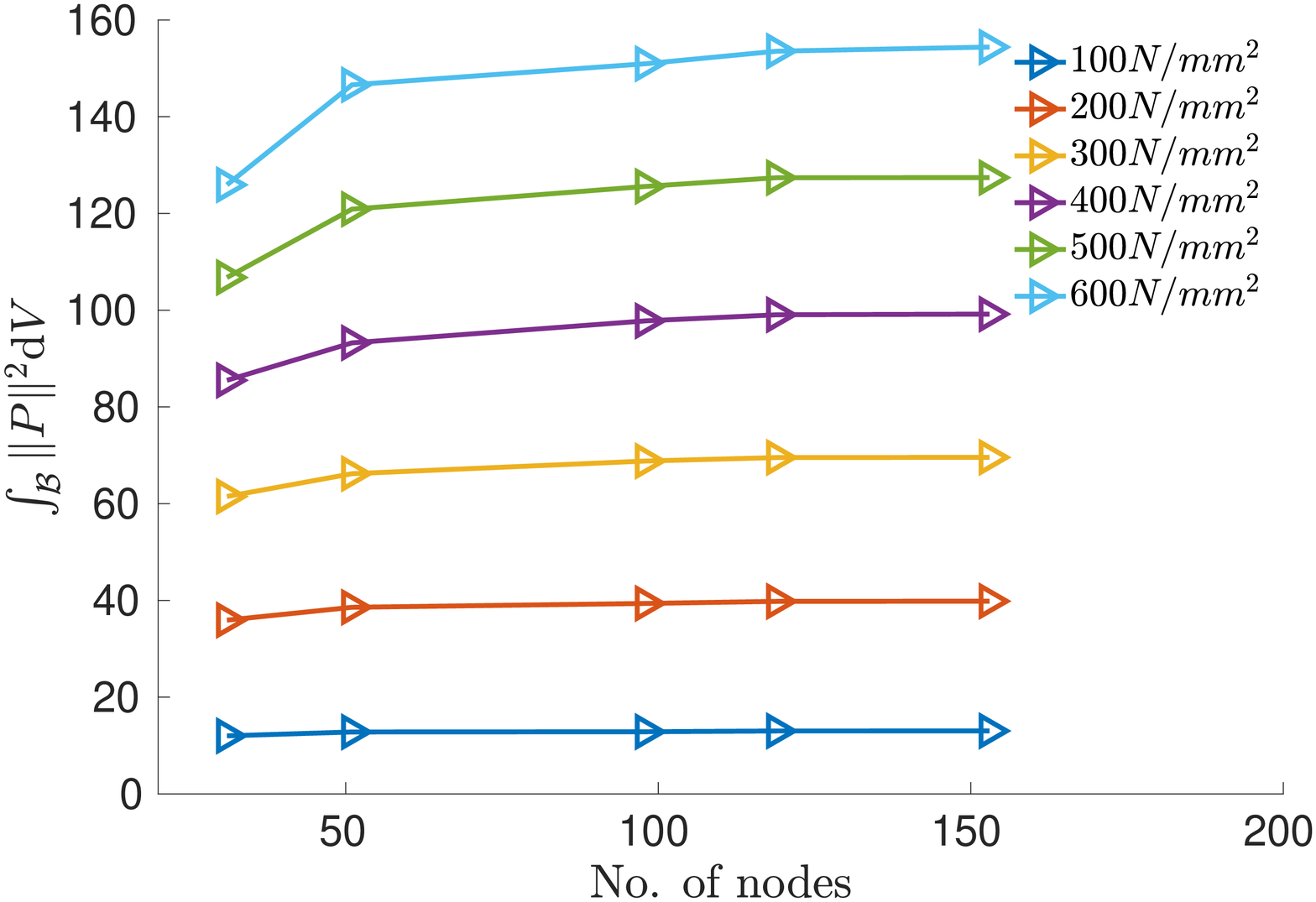}
        \caption{Convergence of deformation 1-forms}
    \end{subfigure}
    \begin{subfigure}[b]{0.49\textwidth}
        \includegraphics[width=\textwidth]{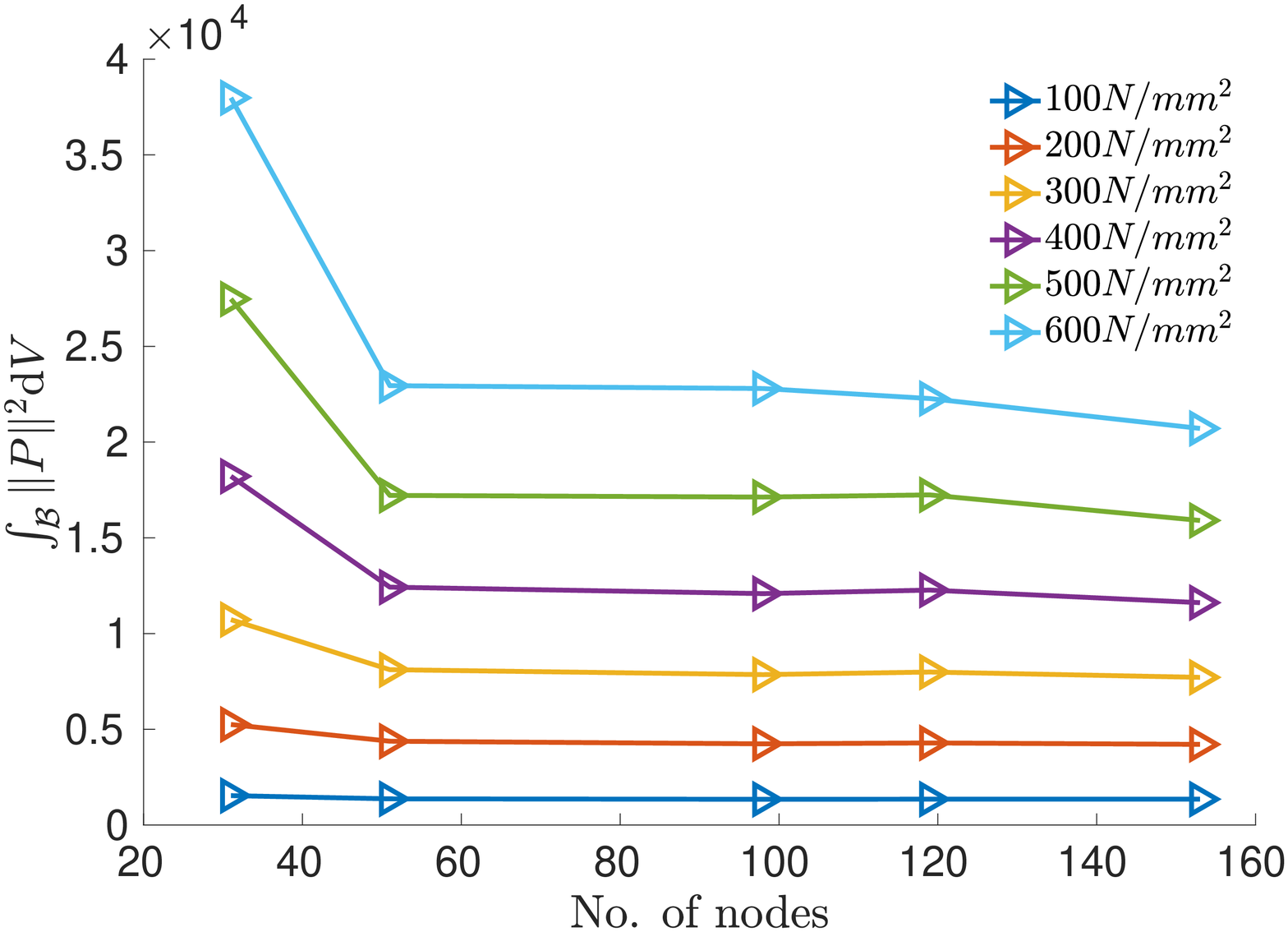}
        \caption{Convergence of first-Piola stress}
    \end{subfigure}
    \caption{Convergence plots for the deformation 1-form and first-Piola stress for different load levels}
    \label{fig:sqCompThetaStressConvergence}
\end{figure}

We now exploit the FE approximation to study the compression of a rectangular block under quasi-incompressible conditions. The neo-Hookian material model with the same material constants as in Cook's membrane problem are used in this simulation. The domain and boundary conditions are shown in Figure \ref{fig:sqCompBvP}.  The convergence of displacement at the point $A$ for different finite element meshes is shown in Figure \ref{fig:sqCompDispConvergence}. For comparison, the displacement convergence obtained using the present FE scheme is reported along with the corresponding plots obtained by Reese \cite{reese2002} and Angoshtari \textit{et al.}\cite{angoshtari2017}. The deformed configuration predicted by our FE formulation is also shown in Figure \ref{fig:DeformedCofigMirror}. The convergence of deformation 1-forms and Piola stress are also computed and presented in \ref{fig:sqCompThetaStressConvergence}. From these convergence curves, it is amply clear that the field quantities of interest converge well even with a relatively coarse mesh.

\subsection{Extension of plate with hole}
\begin{figure}
    \begin{subfigure}[b]{0.49\textwidth}
        \includegraphics[scale=1.0]{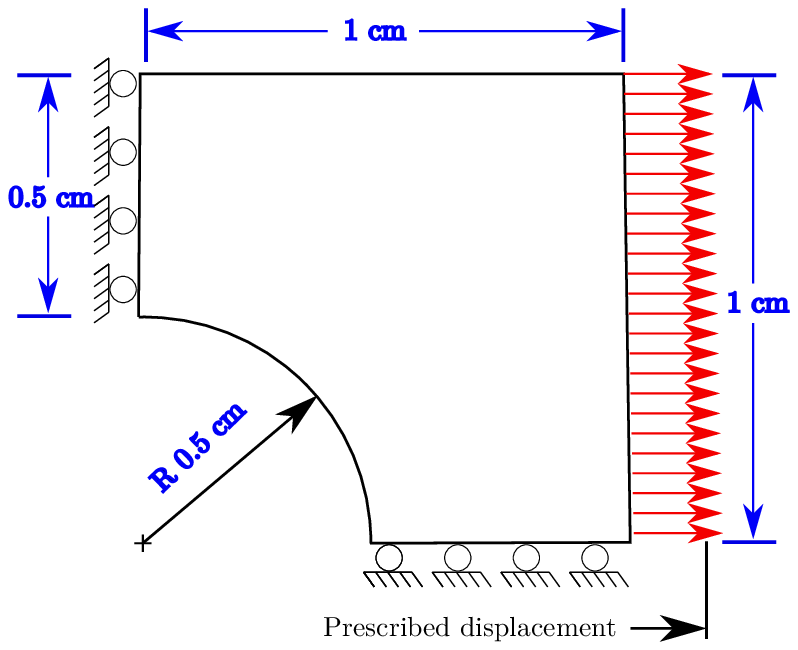}
        \caption{Quarter model of a plate with a hole}
    \end{subfigure}
    \begin{subfigure}[b]{0.49\textwidth}
        \includegraphics[width=\textwidth]{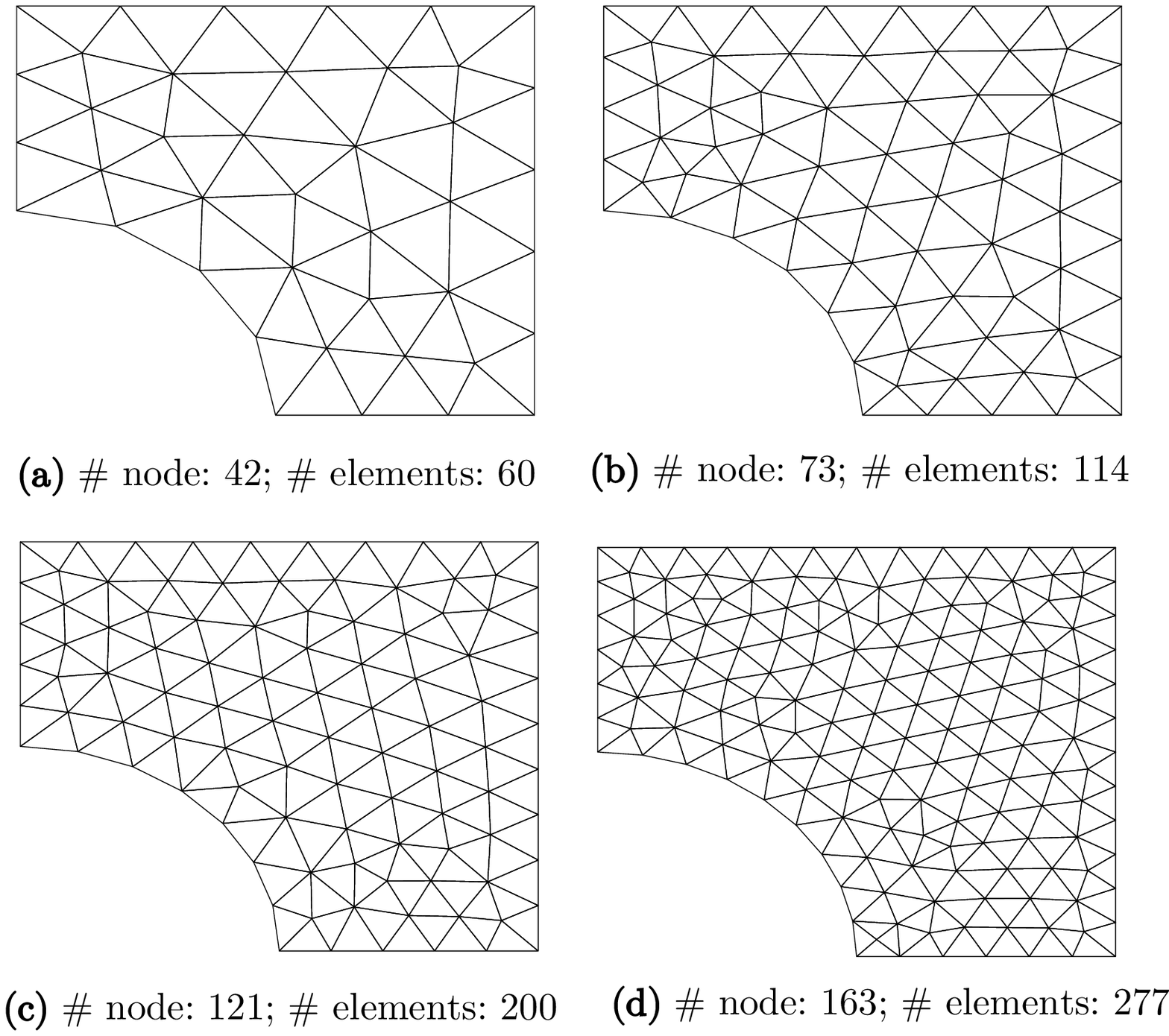}
        \caption{Representative sequence of meshes used in the convergence study}
    \end{subfigure}
    \caption{Boundary conditions and FE mesh used in the convergence study}
    \label{fig:pHoleBvP}
\end{figure}

\begin{figure}
    \begin{subfigure}[b]{0.49\textwidth}
        \includegraphics[width=\textwidth]{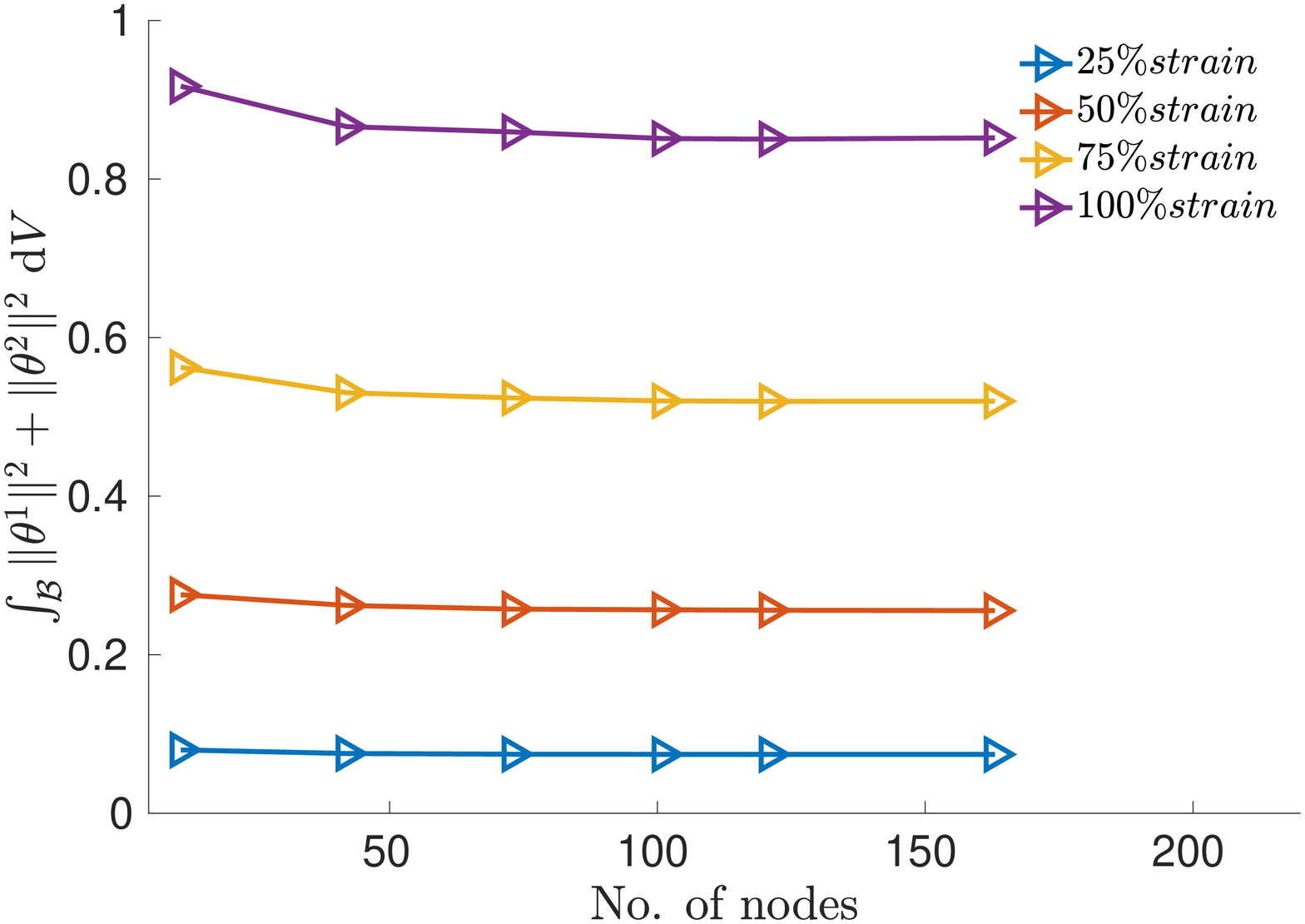}
        \caption{Convergence of deformation 1-forms}
    \end{subfigure}
    \begin{subfigure}[b]{0.49\textwidth}
        \includegraphics[width=\textwidth]{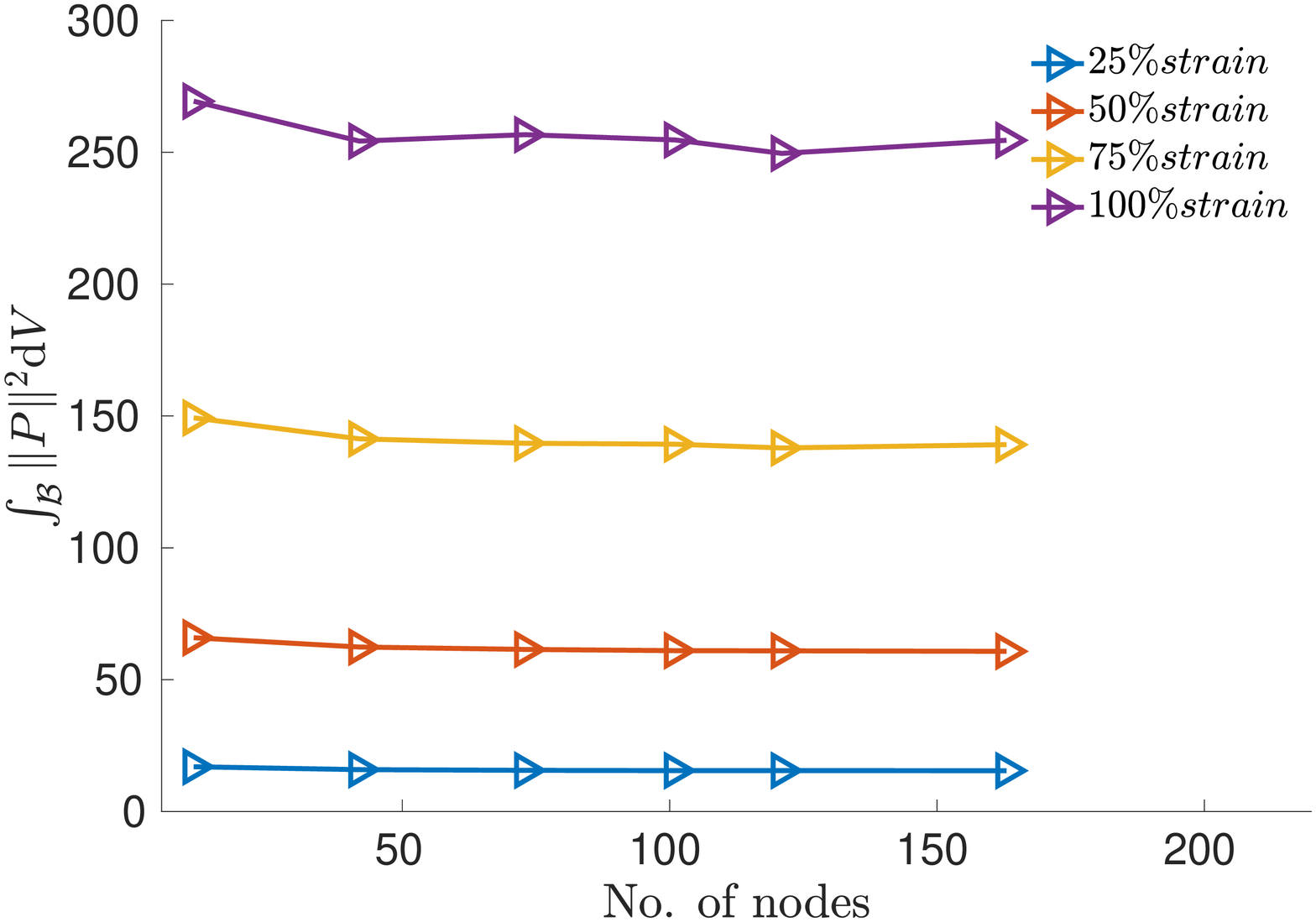}
        \caption{Convergence of first-Piola stress}
    \end{subfigure}
    \caption{Convergence plots of deformation 1-form and first-Piola stress for different load levels}
    \label{fig:pHoleConvergence}
\end{figure}

\begin{figure}
    \centering
    \includegraphics[scale=0.35]{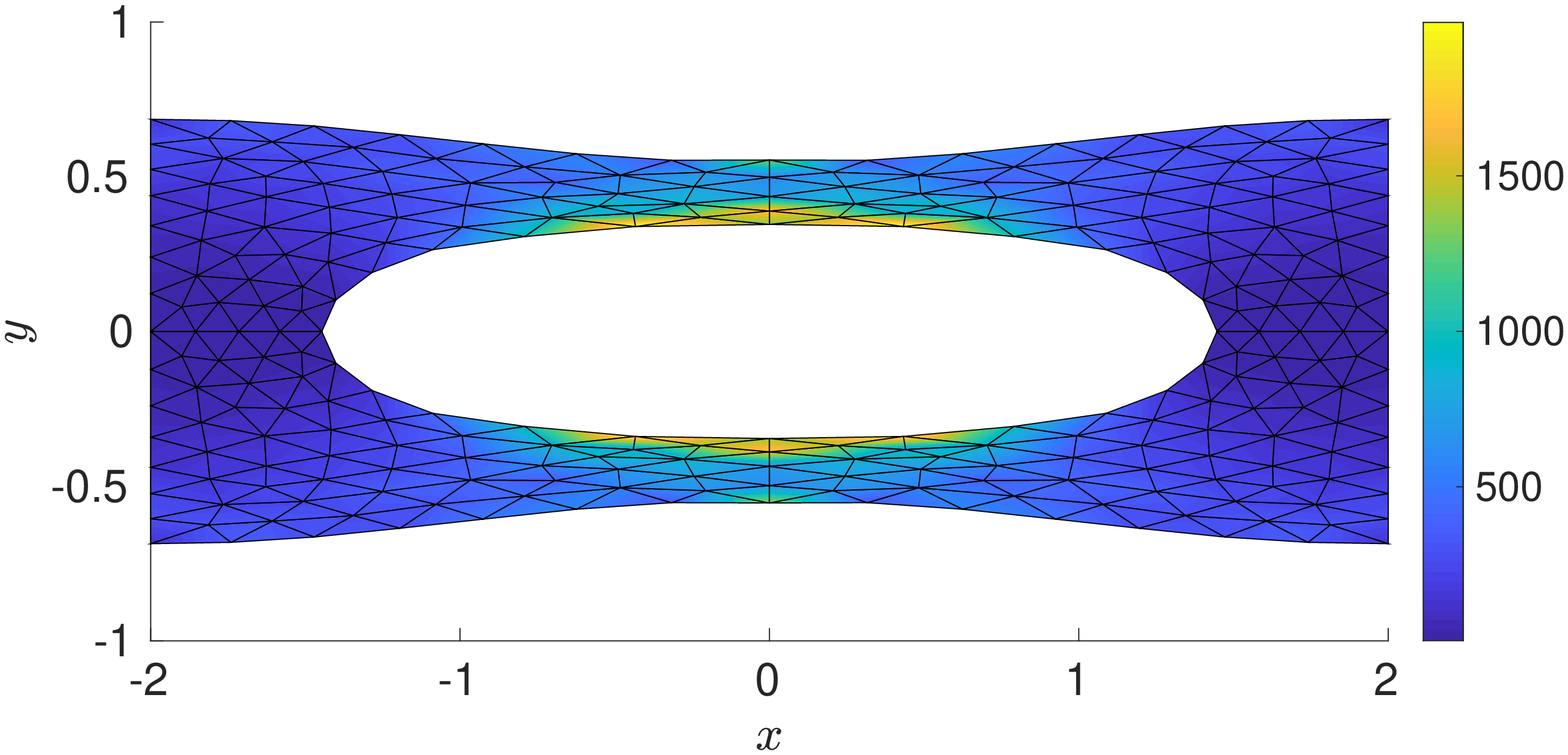}
    \includegraphics[scale=0.35]{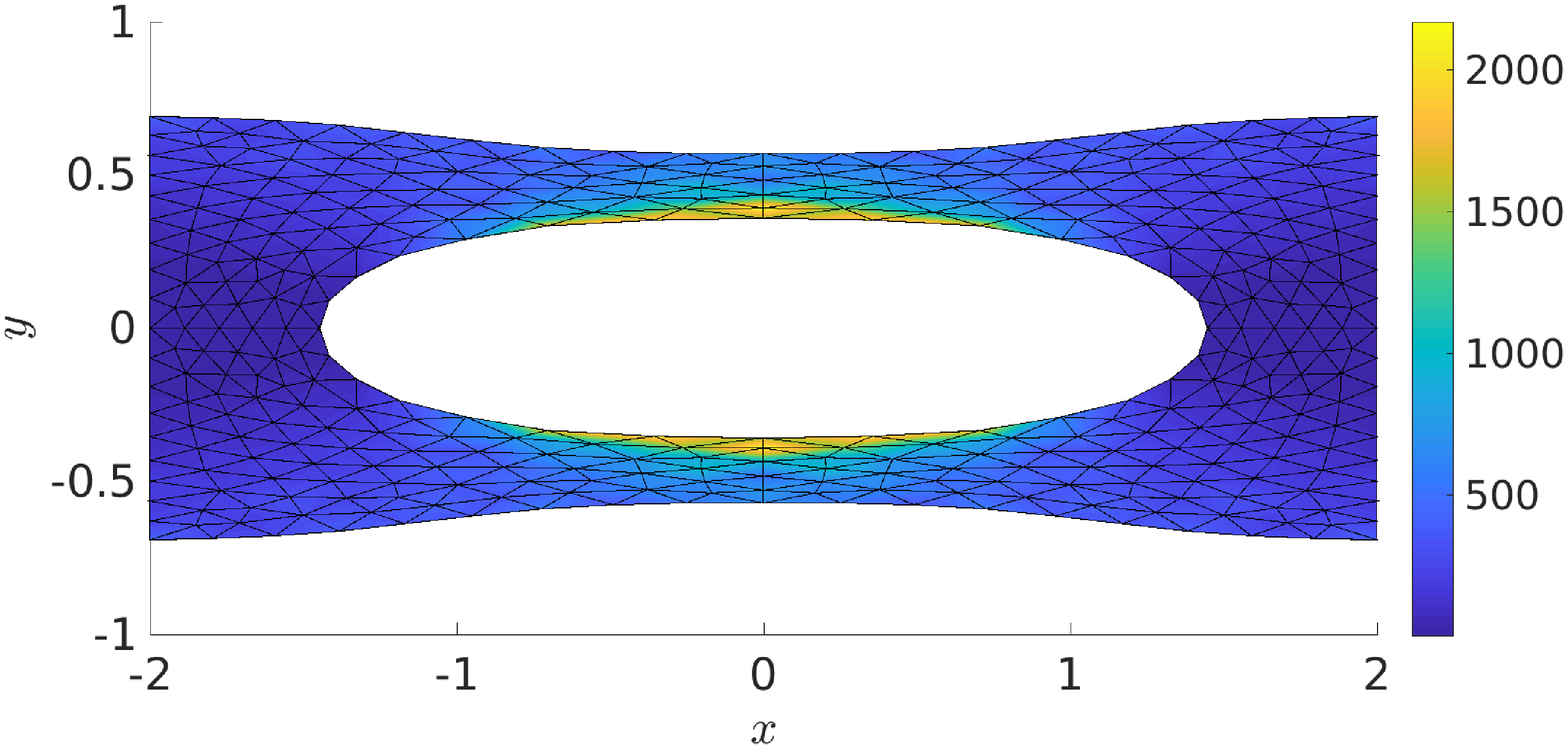}
    \includegraphics[scale=0.35]{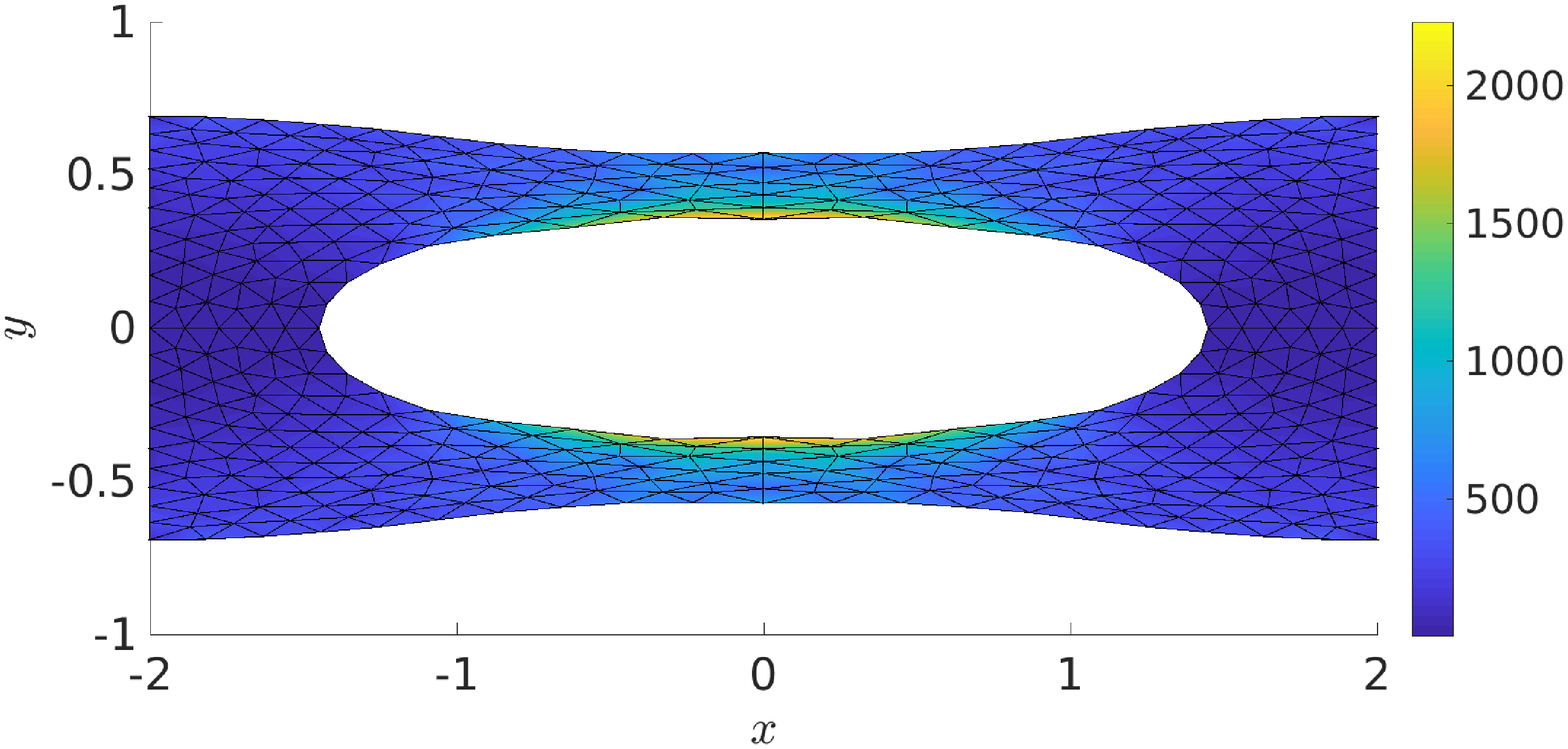}
     \caption{Deformed configuration of the square plate with a circular hole for an extension of 1 $cm$. Color profile indicates the norm of the first Piola stress.}
    \label{fig:pHoleStressContour} 
\end{figure}

In this subsection, we numerically study the extension of a square plate with
a circular hole. The plate has a side length of 2 $cm$ and a hole of radius 0.5
$cm$ placed at the centroid. Neo-Hookian material model \ref{Eq:neoHookianEnergy} with
the material parameters $\lambda=1000$ and $\mu=10$ is used to study the
response of the plate; the assumed material parameters correspond to a
compressible hyperelastic material. The simulation is performed under a
displacement controlled condition, with displacements prescribed along the
face whose normal is along the positive $x-$ direction. Taking into
account the symmetry of the problem, we model only a quarter of the domain.
Boundary conditions used to simulate the deformation of the plate are
presented in Fig. \ref{fig:pHoleBvP}. The convergence of the deformation 1-form and Piola stress is shown in Figure \ref{fig:pHoleConvergence}. The deformed configuration along with the norm of first Piola stress computed using different FE meshes is shown in Figure \ref{fig:pHoleStressContour}. 

\section{Conclusion}\label{sec:conclusion}
This article has reported an entirely new perspective to mixed variational principle in
nonlinear elasticity using Cartan's moving frames. This approach to the mixed
variational principle facilitates a reformulation of the mixed functional in
terms of differential forms. It is also demonstrated that the equations of
mechanical equilibrium, constitutive rule and compatibility could
be obtained as conditions for the critical point of the reformulated mixed
functional. A closer examination of the critical points reveals that in the
absence of compatibility, additional stresses could appear in the body,
without altering the constitutive rule. These additional stresses, though spurious in a strictly Euclidean setup, offer a
pointer to an extension of  the present variational principle to problems
with incompatibilities like dislocations and disclinations. Our present approach
to the kinematics of continua using Cartan's moving frame may play a pivotal
role in such an endeavour, since the setup readily permits the incorporation
of torsion and curvature into the formulation. These aspects of the proposed
variational approach require further study and will be pursued in future. 

Our mixed variational principle based on differential forms is utilized to
construct novel mixed FE techniques to solve 2D problems in nonlinear
elasticity. This FE technique uses finite element exterior calculus to discretise the differential forms
appearing in the variational principle. The present FE method does not require
additional stabilization terms, since it strictly adheres to the algebraic and geometric
structures defined by differential forms, even at the discrete level. The
numerical studies using the our FE technique clearly demonstrate
its superior convergence characteristics.
The extension of the present numerical technique to 3D problems in nonlinear
elasticity is the immediate work in progress, the results of which will be published in a separate article. 

\section*{Acknowledgements}
BD and JK were supported by ISRO through the Centre of Excellence in Advanced
Mechanics of Materials grant No. ISRO/DR/0133. 

\newpage
\bibliographystyle{abbrv}
\bibliography{mainEL}

\begin{thebibliography}{10}

\bibitem{angoshtari2017}
A.~Angoshtari, M.~F. Shojaei, and A.~Yavari.
\newblock Compatible-strain mixed finite element methods for 2d compressible
  nonlinear elasticity.
\newblock {\em Computer Methods in Applied Mechanics and Engineering},
  313:596--631, 2017.

\bibitem{arnold2010}
D.~Arnold, R.~Falk, and R.~Winther.
\newblock Finite element exterior calculus: from hodge theory to numerical
  stability.
\newblock {\em Bulletin of the American mathematical society}, 47(2):281--354,
  2010.

\bibitem{arnold2006}
D.~N. Arnold, R.~S. Falk, and R.~Winther.
\newblock Finite element exterior calculus, homological techniques, and
  applications.
\newblock {\em Acta numerica}, 15:1--155, 2006.

\bibitem{arnold2009geometric}
D.~N. Arnold, R.~S. Falk, and R.~Winther.
\newblock Geometric decompositions and local bases for spaces of finite element
  differential forms.
\newblock {\em Computer Methods in Applied Mechanics and Engineering},
  198(21-26):1660--1672, 2009.

\bibitem{blume1989}
J.~A. Blume.
\newblock Compatibility conditions for a left cauchy-green strain field.
\newblock {\em Journal of elasticity}, 21(3):271--308, 1989.

\bibitem{bossavit1988}
A.~Bossavit.
\newblock Whitney forms: A class of finite elements for three-dimensional
  computations in electromagnetism.
\newblock {\em IEE Proceedings A (Physical Science, Measurement and
  Instrumentation, Management and Education, Reviews)}, 135(8):493--500, 1988.

\bibitem{clelland2017}
J.~N. Clelland.
\newblock {\em From Frenet to Cartan: the method of moving frames}, volume 178.
\newblock American Mathematical Soc., 2017.

\bibitem{Veubeke1974}
B.~F. De~Veubeke.
\newblock Variational principles and the patch test.
\newblock {\em International Journal for Numerical Methods in Engineering},
  8(4):783--801, 1974.

\bibitem{doyle1956nonlinear}
T.~Doyle and J.~L. Ericksen.
\newblock Nonlinear elasticity.
\newblock In {\em Advances in applied mechanics}, volume~4, pages 53--115.
  Elsevier, 1956.

\bibitem{eringen2012}
A.~C. Eringen.
\newblock {\em Microcontinuum field theories: I. Foundations and solids}.
\newblock Springer Science \& Business Media, 2012.

\bibitem{eshelby1951}
J.~D. Eshelby.
\newblock The force on an elastic singularity.
\newblock {\em Philosophical Transactions of the Royal Society of London.
  Series A, Mathematical and Physical Sciences}, 244(877):87--112, 1951.

\bibitem{frankel2011}
T.~Frankel.
\newblock {\em The geometry of physics: an introduction}.
\newblock Cambridge university press, 2011.

\bibitem{guggenheimer1963}
H.~Guggenheimer.
\newblock {\em Differential Geometry}.
\newblock McGraw-Hill series in higher mathematics. McGraw-Hill, 1963.

\bibitem{hehl2012}
F.~W. Hehl and Y.~N. Obukhov.
\newblock {\em Foundations of classical electrodynamics: Charge, flux, and
  metric}, volume~33.
\newblock Springer Science \& Business Media, 2012.

\bibitem{hiptmair1999}
R.~Hiptmair.
\newblock Canonical construction of finite elements.
\newblock {\em Mathematics of computation}, 68(228):1325--1346, 1999.

\bibitem{hiraniThesis}
A.~N. Hirani.
\newblock {\em Discrete exterior calculus}.
\newblock PhD thesis, California Institute of Technology, 2003.

\bibitem{kanso2007}
E.~Kanso, M.~Arroyo, Y.~Tong, A.~Yavari, J.~G. Marsden, and M.~Desbrun.
\newblock On the geometric character of stress in continuum mechanics.
\newblock {\em Zeitschrift f{\"u}r angewandte Mathematik und Physik},
  58(5):843--856, 2007.

\bibitem{mfe}
J.~E. Marsden and T.~J. Hughes.
\newblock {\em Mathematical foundations of elasticity}.
\newblock Courier Corporation, 1994.

\bibitem{marsden2013}
J.~E. Marsden and T.~S. Ratiu.
\newblock {\em Introduction to mechanics and symmetry: a basic exposition of
  classical mechanical systems}, volume~17.
\newblock Springer Science \& Business Media, 2013.

\bibitem{nedelec1980}
J.-C. N{\'e}d{\'e}lec.
\newblock Mixed finite elements in $\mathbb{R}^3$.
\newblock {\em Numerische Mathematik}, 35(3):315--341, 1980.

\bibitem{nedelec1986}
J.-C. N{\'e}d{\'e}lec.
\newblock A new family of mixed finite elements in $\mathbb{R}^3$.
\newblock {\em Numerische Mathematik}, 50(1):57--81, 1986.

\bibitem{oden1974dual}
J.~T. Oden and J.~N. Reddy.
\newblock On dual-complementary variational principles in mathematical physics.
\newblock {\em International Journal of Engineering Science}, 12(1):1--29,
  1974.

\bibitem{ogden1997}
R.~W. Ogden.
\newblock {\em Non-linear elastic deformations}.
\newblock Courier Corporation, 1997.

\bibitem{raviart1977}
P.-A. Raviart and J.-M. Thomas.
\newblock A mixed finite element method for 2-nd order elliptic problems.
\newblock In {\em Mathematical aspects of finite element methods}, pages
  292--315. Springer, 1977.

\bibitem{reese2002}
S.~Reese.
\newblock On the equivalent of mixed element formulations and the concept of
  reduced integration in large deformation problems.
\newblock {\em International Journal of Nonlinear Sciences and Numerical
  Simulation}, 3(1):1--34, 2002.

\bibitem{reese2000}
S.~Reese and P.~Wriggers.
\newblock A stabilization technique to avoid hourglassing in finite elasticity.
\newblock {\em International Journal for Numerical Methods in Engineering},
  48(1):79--109, 2000.

\bibitem{rognes2010}
M.~E. Rognes, R.~C. Kirby, and A.~Logg.
\newblock Efficient assembly of h(div) and h(curl) conforming finite elements.
\newblock {\em SIAM Journal on Scientific Computing}, 31(6):4130--4151, 2010.

\bibitem{segev1999}
R.~Segev and G.~Rodnay.
\newblock Cauchy's theorem on manifolds.
\newblock {\em Journal of Elasticity}, 56(2):129--144, 1999.

\bibitem{shojaei2018}
M.~F. Shojaei and A.~Yavari.
\newblock Compatible-strain mixed finite element methods for incompressible
  nonlinear elasticity.
\newblock {\em Journal of Computational Physics}, 361:247--279, 2018.

\bibitem{shojaei2019}
M.~F. Shojaei and A.~Yavari.
\newblock Compatible-strain mixed finite element methods for 3d compressible
  and incompressible nonlinear elasticity.
\newblock {\em Computer Methods in Applied Mechanics and Engineering},
  357:112610, 2019.

\bibitem{simo1989}
J.~C. Simo and D.~D. Fox.
\newblock On a stress resultant geometrically exact shell model. i: Formulation
  and optimal parametrization.
\newblock {\em Computer Methods in Applied Mechanics and Engineering},
  72(3):267--304, 1989.

\bibitem{simo1986}
J.~C. Simo and T.~J.~R. Hughes.
\newblock On the variational foundations of assumed strain methods.
\newblock {\em Journal of Applied Mechanics}, 53(1):51--54, 1986.

\bibitem{whitney2012}
H.~Whitney.
\newblock {\em Geometric integration theory}.
\newblock Courier Corporation, 2012.

\bibitem{yavari2008}
A.~Yavari.
\newblock On geometric discretization of elasticity.
\newblock {\em Journal of Mathematical Physics}, 49(2):022901, 2008.

\bibitem{yavari2013}
A.~Yavari.
\newblock Compatibility equations of nonlinear elasticity for
  non-simply-connected bodies.
\newblock {\em Archive for Rational Mechanics and Analysis}, 209(1):237--253,
  2013.

\end{thebibliography}

\end{document}